\documentclass[singlespacing]{elsart}
\usepackage{graphicx}
\usepackage{amsmath}
\usepackage{amsfonts}
\usepackage{amsbsy}
\usepackage{amssymb}
\usepackage{setspace}
\usepackage{bm}
\journal{Icarus}

\long\def\symbolfootnote[#1]#2{\begingroup%
\def\thefootnote{\fnsymbol{footnote}}\footnote[#1]{#2}\endgroup}

\renewcommand{\u}{\ensuremath{\mathbf{u}}}
\renewcommand{\d}{\ensuremath{\partial}}

\newcommand{\ey}{\ensuremath{\mathbf{e}_{y}}}

\newcommand\p{\partial}
\newcommand\f{\frac}

\newcommand\twothirds{{\textstyle\frac{2}{3}}}
\newcommand\fourthirds{{\textstyle\frac{4}{3}}}
\newcommand\seventhirds{{\textstyle\frac{7}{3}}}

\newcommand\rmD{\mathrm{D}}

\begin{document}

\begin{frontmatter}

\title{The viscous overstability, nonlinear wavetrains, and finescale
  structure in dense planetary rings }
\author[cam1,cam2]{Henrik N. Latter\corauthref{cor}},
\corauth[cor]{Corresponding author.}
\author[cam2]{Gordon I. Ogilvie}
\address[cam1]{Address: LRA, 
 \'{E}cole Normale Sup\'{e}rieure, 24 rue Lhomond, Paris 75005, France.
   Email: henrik.latter@lra.ens.fr}
\address[cam2]{Address: DAMTP, University of Cambridge, Wilberforce Rd, Cambridge CB3
  0WA, United Kingdom.
Email: gio10@cam.ac.uk}

\begin{abstract}
This paper addresses the fine-scale axisymmetric structure exhibited in 
Saturn's A and B-rings. We aim to explain 
both the periodic microstructure on 150--220m, revealed by the Cassini UVIS and
RSS instruments, and the
irregular variations in brightness on 1--10km, reported by the Cassini ISS. 
We propose
 that the former structures correspond to the peaks and troughs
of the nonlinear wavetrains that form naturally in 
a viscously overstable disk. The latter variations on longer scales
may correspond to modulations and defects in the wavetrains'
amplitudes and wavelength. We explore these ideas using a simple hydrodynamical
model which captures the correct qualitative behaviour of a disk of
inelastically colliding particles, while also permitting us to make progress
 with analytic and
semi-analytic techniques. Specifically, we calculate a family of travelling 
nonlinear
density 
waves and determine their stability properties.
 Detailed numerical simulations that confirm our
basic results will appear in a following paper.

\end{abstract}
\begin{keyword} 
Planetary Rings; Saturn, Rings; Collisional Physics
\end{keyword}

\end{frontmatter}

\section{Introduction}

Saturn's A and B-rings sport an abundance of
irregular radial structure which, though aesthetically pleasing, presents
something of a puzzle to the theoretician.  
The instruments aboard the \emph{Cassini} space probe show that these
patterns manifest on a vast range of length-scales and can take quite
 different forms. For instance, there exist quasi-periodic
microstructure
on scales of 0.1 km (Colwell et al.~2007; Thomson et al.~2007),
discontinuous and irregular striations on the 1-10 km intermediate
scale, 
and much
broader 100 km undulations (Porco et
al.~2005). In addition to the difficulties involved in tackling these three orders
of magnitude, there are 
  the formidable
modelling questions posed by a cold disk of densely-packed, inelastic, and
infrequently colliding particles (Stewart et al.~1984, Araki and Tremaine
1986, Salo 1991, H\"{a}meen-Antilla and
Salo 1993, Schmidt et al.~2001, Latter and Ogilvie 2008). This paper will
focus on only a subset of these phenomena, the smaller-scale variations,
 and will not linger especially on the modelling issues. Specifically, 
it investigates how 
the quasi-periodic microstructure
 relates to the viscous overstability, on one hand,
and to structure formation on the intermediate scales, on the other.
 The variations on the much longer 100km scale are not examined, and we suspect
 that they have their origin in a different mechanism
entirely, perhaps ballistic transport (Durisen 1995).

Our starting point is the viscous overstability, which is
now regarded as a key player in the short scale radial dynamics
 of Saturn's rings (Schmit and Tscharnuter
1995, Schmidt et al.~2001, Spahn and Schmidt 2006, Latter and Ogilvie
2008). The viscous overstability is an axisymmetric oscillatory instability
that afflicts the homogeneous
state of Keplerian shear. Growing modes rely on the
 alliance of the fluid disk's inertial-acoustic
 oscillations
 with the disk's stress oscillations: variations in the stress extract energy from the
Keplerian shear and inject it into the inertial-acoustic wave; but the increased
 motion this induces magnifies the stress oscillation itself which can extract even more
energy, and so the process runs away. The feedback loop requires (a) the stress to
efficiently remove energy from the Keplerian flow and (b) the two oscillations
to communicate effectively, in particular for them to be in phase. The first
 condition is tied to the stress's sensitivity to surface density. The second
 condition is often violated in dilute
rings and turbulent disks, where the stress can lag behind the epicycles 
and the overstability fails to work
 (Ogilvie 2001, Latter and Ogilvie 2006a).

 The instability's linear theory has been well
 established by a variety of theoretical approaches: hydrodynamics (Schmit
 and Tscharnuter 1995, Schmidt et al.~2001), $N$-body simulations (Salo 2001, Salo et
 al. 2001), and kinetics (Latter and Ogilvie 2006a, 2008). However, its
 nonlinear theory has received surprisingly little attention. Two hydrodynamical
 studies exist, a large-scale nonlinear simulation of a ring annulus in 1D  (Schmit and
 Tscharnuter 1999) and a weakly nonlinear analysis (Schmidt and Salo
 2003). The simulations show that the nonlinear evolution of an overstable disk
 is characterised by significant disorder.
In contrast, the latter study
  suggests that an overstable ring may exhibit simple
 coherent structures which take the form of
 travelling waves. Because Schmit and Tscharnuter impose reflecting
 boundary conditions, and hence break translational symmetry,
  such coherent structures may have been difficult to observe in their simulations.
 Certainly, more simulations
  need to be undertaken and the influence of the boundary conditions
 better understood. Concurrently, a fully nonlinear theory
extending the work of Schmidt and Salo is required to fully establish the
 existence of the nonlinear solutions.
  The former project we present elsewhere,
 the latter we present here.

First, we demonstrate that
 steady nonlinear travelling wavetrains are exact solutions to the governing
 nonlinear equations of a viscously overstable disk, and
second, 
that these solutions are the loci of a rich secondary set of dynamics which
generate irregular variations in the waves' amplitude and wavenumber. 
 We propose that viscously overstable regions in the A and B-rings
support a bed of travelling wavetrain solutions and 
these structures correspond to the
 quasi-periodic variations registered by Cassini's UVIS and RSS instruments
(the radial `microstructure'). 
Modulations and defects in the amplitudes and wavelengths of these wavetrains
 may
yield different optical properties which can be traced by the Cassini
cameras. Consequently, we hypothesise that these larger-scale variations
are associated with the intermediate 1-10 km structures observed (Porco et
al. 2005). 
Our hypotheses are investigated
 with a one-dimensional, isothermal fluid
model endowed with a Navier-Stokes stress. Though simplistic,
 it should predict qualitatively correct behaviour, while
permitting the problem to be attacked analytically or
semi-analytically. Importantly, self-gravity is omitted throughout, but
we investigate its role in a following paper. This is mainly for readability
as the nonself-gravitating dynamics is sufficiently complex on its own.

  A study of nonlinear waves connects
  naturally to the extensive field of wave propagation in thin
  astrophysical disks (Kato 2000), and 
in particular to the launching of spiral
  density waves (Goldreich and Tremaine 1978a, 1979, 1980,
  Borderies et al.~1983, 1986, Shu
  et al.~1985a, 1985b).
In the latter studies
  the emphasis has been on the launching of spiral waves by an external
  potential, as might issue from a moon, and the damping of such waves
   by a viscous stress. But the viscous stress can be
 an engine of growth also, as in the development of
  global eccentric modes and narrow ringlets (Borderies et al.~1985, 
Papaloizou and Lin 1988, Lyubarskij et al.~1994, Longaretti and Rappaport
  1995, Ogilvie 2001) and the axisymmetric
  viscous overstability itself (Kato 1978, Schmit and Tscharnuter 1995). In this paper 
  it will be shown that the viscous forces in a planetary ring can balance
  energy dissipation and injection and thus sustain
  \emph{steady} wave structures.

 A study of nonlinear wavetrains also connects to the
  perhaps more massive field of nonlinear waves in general media and 1D reaction--diffusion
  systems especially, the model equation of which is the complex
  Ginzburg-Landau equation (Aranson and Kramer 2002).
 The dynamics of the complex Ginzburg-Landau equation
 is remarkably rich and may provide a template for the study of viscous
  overstability in Saturn's rings. As in a fluid disk, the equation admits a trivial
  homogeneous solution that is susceptible to an oscillatory linear
  instability via a Hopf bifurcation
 (the analogue of the viscous overstability); it also supports both stable and
  unstable steady nonlinear wavetrains. Variations upon the wavetrains
  exhibit a wide variety of behaviours ranging from smooth modulations to
  abrupt jumps in wavenumber and amplitude (sources and shocks), as well as
   small-scale chaotic fluctuations (Bernoff 1988, Shraiman et al.~1992,
  Chat\'{e} 1994,
  Aranson and Kramer 2002).
 In particular, the sources and shocks
can partition the radial domain into a one-dimensional cellular structure, in
  which each `cell' is characterised by a different wavetrain (Chat\'{e}
  1993, Popp et al.~1994, Ipsen and Hecke 2001). The latter behaviour seems
  especially pertinent to ring structure on intermediate scales, which
  Cassini's cameras
  show as a pattern of irregularly spaced, disjunct bands (Porco et al.~2005).

\vskip0.3cm

A summary of the paper is as follows. First, space will be lent to
a discussion of the validity of the simple hydrodynamic
model that we employ, the parameters it introduces, and the values these
should take. In Section 3,
which comes next, a brief summary of the linear stability of the homogeneous
state of Keplerian shear will be presented. Though these results have appeared
a number of times elsewhere (Schmit and Tscharnuter 1995, Schmidt et
al. 2001), we include them for completeness. Section 4 will demonstrate the
existence and investigate the properties of axisymmetric, steady, nonlinear, travelling
wavetrains in a viscously overstable disk. The nonlinear wave profiles
are calculated numerically via the solution of a nonlinear eigenvalue problem
and analytically in the limit of long wavelength. We find that an overstable
disk supports a one-parameter family of wavetrain solutions, the members
of which can be parametrised by
their wavenumber $\lambda$.
In Section 5 we establish the linear stability of the wavetrains 
 and find that above a critical wavelength $\lambda_\text{st}$ the wavetrains
are linearly stable. For model parameter values, associated with
optical depths of 1--1.5, the critical wavelength is approximately $50H$,
where $H$ is the scale height of the disk. The critical wavelength appears
 consistent with Cassini observations of periodic microstructure. 
Lastly, in Section 6, we discuss more fully the
general nonlinear dynamics of an overstable disk. A rough argument is sketched
 explaining the upward cascade to longer scales observed in
 the simulations of
 Schmit and Tscharnuter (1999) and we suggest that,
irrespective of self-gravity, this `inverse cascade' should halt (or at least change
qualitatively) once
 most power reaches a
lengthscale near $\lambda_\text{st}$. A discussion
follows which addresses what this saturated state may look like.
Modulations and defects in the wavetrains'
wavenumber and amplitude are touched on, and a multiple-scales
analysis presented in which we show that sources and shocks in the wavetrains'
phase may develop. Finally we discuss the relevance of the complex 
Ginzburg-Landau equation.
Section 7 presents our conclusions, in which we summarise the paper, and point
out the issues which remain unresolved and necessary future work.

\section{The model}

In order to bring out the important features of the ring's nonlinear
evolution we employ a simple hydrodynamical model that captures
 the correct qualitative behaviour while not burdening the analysis
or obscuring its interpretation with mathematical complexity.
 The ring is assumed to be a
vertically-averaged, non-self-gravitating,
 isothermal, ideal gas possessing a Newtonian viscous stress. In addition, the shearing and
 bulk viscosities depend on surface density as power
 laws, in accordance with Schmit and Tscharnuter (1995, 1999) and Schmidt et
 al.~(2001). We also work with the shearing sheet approximation,
 which is a Cartesian representation of
 a `patch' of disk orbiting the central planet with frequency $\Omega$, and in which
 $x$ and $y$ denote the radial and azimuthal dimensions respectively (see
 Goldreich and Lynden-Bell 1965).

 The governing equations are
 \begin{align} \label{sigmad}
\p_t\sigma+\p_i(\sigma u_i)&=0, \\
  \sigma(\p_tu_i+u_j\p_ju_i+2\Omega\epsilon_{izj}u_j)&=-\sigma\p_i\Phi_T-\p_iP+
  \p_j\Pi_{ij}, \label{ud}
\end{align}
where $\sigma$, $P$, and $\Pi_{ij}$ are the vertically integrated
density, pressure, and viscous stress, and $u_i$ is the planar fluid
velocity. The tidal potential is
$\Phi_T=-3\Omega^2 x^2/2$.

The pressure comes from the ideal gas equation of state
\begin{equation}
P= v_s^2 \sigma,
\end{equation}
where $v_s$ is the isothermal sound speed.
The viscous stress is
given by
\begin{equation} \label{Pi}
  \Pi_{ij}=\nu\sigma(\p_i u_j+\p_j u_i)+
  (\nu_b-\twothirds\nu)\sigma(\p_k u_k)\delta_{ij}.
\end{equation}
The kinematic shear and bulk viscosities are parametrised as
\begin{align} \label{nupresc}
\nu= \frac{v_s^2}{\Omega}\,\alpha\left(\frac{\sigma}{\sigma_*}\right)^\beta, \qquad
\nu_b= \frac{v_s^2}{\Omega}\,\alpha_b\left(\frac{\sigma}{\sigma_*}\right)^\beta,
\end{align}
where $\alpha$, $\alpha_b$, and $\beta$ are dimensionless parameters,
 and $\sigma_*$ is the surface density of the
 homogeneous state of Keplerian shear.

Throughout the paper we employ dimensions so that
$$  \Omega= 1,  \quad v_s=1,  \quad
\sigma_*=1. $$
 One unit of time 
is therefore $(2\pi)^{-1}$ times an orbital period and the unit of length
is $H\equiv v_s/\Omega$, the scale height of the disk.
 The
full set of governing parameters is now $\alpha$, $\alpha_b$, and
$\beta$.

\subsection{Assumptions}

Of course,  much can be
 said about each of the model assumptions. We shall say but a little.
First, the adoption of vertical averaging limits
 us to radial scales much longer than the disk scale height, $H$.
  The shearing sheet on the other hand
 limits us to scales much shorter than the disk radius. Neither
 constraint is much of a problem 
because the lengthscales of the phenomena we examine sit well within this
 enormous range. More of an issue is the neglect of the vertical motions.
 For instance, the viscous
 overstability is usually accompanied by vertical `breathing' or
 `splashing' and may be significant when combined with nonisothermal
 behaviour. We, however, leave this issue open for the time being.

A real planetary ring is not generally isothermal. Due to the
 relative infrequency of collisions, the thermal
 time-scale is of the same order as the dynamical time-scale, as
 kinetic theories and $N$-body simulations show (Goldreich and
 Tremaine 1978b, Stewart et al.~1984, Brahic 1977).
 That said, when collisions are a little more frequent ---
as in a dense ring which may support some tens
 of collisions per orbit --- the isothermal model affords
an acceptable approximation (Salo et
al. 2001, Latter and Ogilvie 2008). 

A dense ring is not an ideal gas, and is probably
  better suited to a polytropic equation of state which can better
 capture the
 effects of close packing and nonlocal pressure (Schmidt et al.~2001,
 Schmidt and Salo 2003, Latter and Ogilvie 2008). We persist, however,
 with the simpler model in the belief that the errors introduced do
 not alter the qualitative behaviour.

Finally, it should be acknowledged
 that the viscous stress of a planetary ring behaves in a way
  not always captured by a simple 
Navier-Stokes stress. For instance, in a dilute ring, effects associated with
nonlocality in time (`memory effects') are important 
because the translational stress relaxes on a time-scale comparable to the
dynamical time-scale (Latter and Ogilvie 2006a). In contrast, a dense
ring possesses a stress dominated by the
collisional component, and, though local in time, will be
 a nonlinear
function of the rate of strain (Latter and Ogilvie 2008).
The Navier-Stokes model, however, offers a reasonable approximation of the dense system,
particularly as we may `tweak' the parameter $\beta$ in order to roughly
 `compensate' for the linearisation in strain.

Self-gravity is excluded in this paper and will be examined
separately in a subsequent article. It undoubtedly plays a role in both the
axisymmetric and nonaxisymmetric dynamics of an overstable disk
 but the effects are complicated
and deserve a special treatment.
In addition, $N$-body simulations show that nonaxisymmetric self-gravity wakes
 hinder the development
of the linear modes (Schmidt et al.~2001), and it is most probable that they impact on the nonlinear dynamics as
well. This is an important issue but one that can only be resolved
satisfactorily by three-dimensional simulations.

\subsection{Parameters}

Aside from these theoretical issues, we are encouraged to use the
isothermal hydrodynamic model because it predicts qualitatively correct
behaviour when compared with
$N$-body simulations. The linear growth rates of the overstable modes
are adequately approximated (Schmidt et al.~2001) as is the weakly
nonlinear evolution of their amplitudes (Schmidt and Salo 2003).
For these reasons the
hydrodynamic parameter values we use will be set equal to those computed
 in these studies, in particular, from Salo et al.'s (2001) $N$-body simulations,
which treated
particles of radius $100$ cm, undergoing collisions according to the
Bridges et al.~(1984) piecewise power law at a radius of $100,000$ km.
 Full self-gravity was not
included in the simulations, but its compression of the disk thickness was mimicked by
increasing the vertical epicyclic frequency of the particles
$\Omega_z$ (an idea pioneered by Wisdom and Tremaine 1988).
 In Table 1 we reproduce some of the data of these runs
for different optical depth $\tau$ and with a vertical frequency
enhancement of $\Omega_z/\Omega=3.6$.

These values will serve us only as a guide.
The viscous parameters are closely tied to the
 kinetic parameters of a particular simulation
 (such as particle size, elasticity law)
 but their complete functional dependence is far from understood. On the other
 hand,
 the kinetic parameters
of a real particulate ring are not yet fully constrained. Thus we
let $\alpha$, $\alpha_b$ and especially $\beta$ take a variety of values.
 The $\beta$ parameter is the quantity most sensitive to the background optical
depth, as we can see from Table 1, and also to self-gravity. With respect to
the last point, Schmidt and Salo
(2003) have published another set of parameter values from a simulation in
which $\Omega_z/\Omega=2$, not 3.6, and they find that $\beta$ takes
significantly smaller values in this case. On the other hand, the
simplifying assumption of a Newtonian stress may require us to vary
$\beta$; and indeed, Schmidt and Salo (2003) inflate $\beta$ by up to
$13\%$ in order to obtain agreement in their weakly
nonlinear analysis. Finally, in a real disk, the existence of
gravitational wakes (screened out by the $N$-body simulations we mention)
 will militate against the
development of the overstable modes, meaning that $\beta$ may take
a smaller `effective' value. The actual situation, of course, is probably more
complicated,
not least by the additional effects of nonlocal `gravitational viscosity'
(Daisaka et al.~2001).

\begin{table}[!ht]
\begin{center}
\begin{footnotesize}
\begin{tabular}{|c|c|c|c|c|c|}
\hline 
 $\tau$ & $\alpha$ & $\alpha_b$ &
 $\beta$& $v_s/\Omega$ (m)  \\ 
\hline
0.5& 0.348&1.08&0.67&2.47\\
\hline
1.0&0.357& 0.764&1.15&3.29 \\
\hline
1.5&0.342&0.681&1.19&4.42\\
\hline
2.0&0.322 &0.683&1.55&5.45\\
\hline
\end{tabular}
 \end{footnotesize}
\end{center}
  \caption{\footnotesize{Nondimensionalised hydrodynamical parameters at different optical
 depths $\tau$ derived from $N$-body simulations
 for a disk of $100$~cm radius particles, at a radius of
 $100,000$~km, undergoing collisions according to the Bridges et
 al.~(1984) elasticity law, with vertical frequency enhancement of
 $\Omega_z/\Omega=3.6$ (Schmidt et al.~2001, Salo et al.~2001).}}
\end{table}

\section{Linear stability of the homogeneous steady state}

This section presents a summary of
 the stability analysis of the homogeneous steady state of
Keplerian shear. Though it has been thoroughly examined
 with various continuum models
( Lin and Bodenheimer 1981, Ward 1981, Stewart et
al.~1984, Schmit and
Tscharnuter 1995, Spahn et al.~2000, Schmidt et al.~2001, Latter and
Ogilvie 2006a, 2008) we repeat the analysis for completeness and to connect it
 to our nonlinear work in Sections 3 and 4.

\begin{table}[ht!]
\begin{center}
\begin{footnotesize}
\begin{tabular}{|c|c|c|}
\hline
\textbf{Symbol} & \textbf{Definition} \\
\hline
$\tau$ & Normal geometric optical depth \\
\hline 
 $\sigma$ & Surface density  \\ 
\hline
$ \mathbf{u}$, $\mathbf{u}'$ & Total and perturbation velocities  \\ 
\hline
 $\Phi_T$  & Tidal potential  \\ 
\hline
$P$ & Pressure \\
\hline
$\mathbf{\Pi}$ & Viscous stress \\
\hline
$f$, $g$ & Dimensionless $xx$ and $xy$ component of the pressure tensor\\
\hline
$\nu$, $\nu_b$  & Shearing and bulk kinematic viscosities  \\
\hline
$\alpha$, $\alpha_b$ & Shearing and bulk `alpha' parameters \\
\hline
$\beta$ & Exponent of density dependence of the kinematic viscosities \\
\hline
$ v_s$ & Sound speed \\
\hline
$\Omega$ & Orbital frequency \\
\hline
$H = v_s/\Omega$ & Disk scale height \\
\hline
$k$, $s$ & Wavenumber and growth rate of a linear disturbance \\
\hline
$\mu$, $\omega$ & Wavenumber and frequency of a nonlinear wavetrain
 \\
\hline
$c_p = \omega/\mu$ & Phase speed of a nonlinear wavetrain \\
\hline
$c_g = d\omega/d\mu$ & Group speed of a nonlinear wavetrain \\
\hline
$\theta = \mu x- \omega t$ & Phase variable of a nonlinear wavetrain \\
\hline
$\lambda=2\pi/\mu$ & Wavelength of a nonlinear wavetrain \\
\hline
$\lambda_\text{st}$ & Critical wavelength above which nonlinear wavetrains are linearly
 stable \\
\hline
$\mathbf{Z}=(\sigma,u_x',u_y')$ & Solution vector \\
\hline
$u=c_p-u_x'$, $v=u_y'$ & Dimensionless velocity perturbation variables \\
\hline
$q$,
$\vartheta(x,t)$ & Amplitude and phase variables for asymptotic long wavelength wavetrains \\
\hline
$\delta$ & Small ordering parameter \\
\hline
$X = \delta(x- c_g t) $, $T=\delta^2 t$ & Long spatial and temporal variables \\ 
\hline
$\Theta(X,T)$, $K= \Theta_X$ & Slowly varying phase and wavenumber of a
nonlinear wavetrain \\
\hline
$c_\text{sh}$ & Speed of a travelling shockfront \\
\hline
\end{tabular}
 \end{footnotesize}
\end{center}
  \caption{\footnotesize{Table of parameters, variables, and symbols}}
\end{table}

The equations \eqref{sigmad} and \eqref{ud} admit the following equilibrium:
 $\sigma=1$ and $u_y= -(3/2)\,x$, $u_x=0$.
We introduce a small perturbation so that
$\sigma=1+\sigma'$ and $\u=-(3/2)\,x\,\ey+\u'$. Their
linearised equations read
\begin{align}\label{hel1}
&\d_t \sigma' + \d_x u_x'=0, \\
&\d_t u_x' - 2u_y' = -\d_x f',\\
&\d_t u_y' + \tfrac{1}{2}u_x' = -\d_x h',
\end{align}
with
\begin{align}
 f'= \sigma'-(\alpha_b+\tfrac{4}{3}\alpha)\d_x u_x', \qquad
 h'= \frac{3}{2}\alpha(1+\beta)\sigma'-\alpha\,\d_x u_y'.\label{hel2}
\end{align}
(For a full list of symbols see Table 2.)
The first term in the expression for $h'$ is the key to the disk's stability.
 It couples the variations in
 the viscous stress with the background shear and thus allows
 energy to be drawn from
 the shear and directed into the perturbation.
 We now represent the perturbation as a Fourier mode  $\propto e^{i k x + s
t}$ where $k$ is the (real) wavenumber and $s$ is the (complex) growth
rate. 
The ansatz is substituted
into Eqs \eqref{hel1}-\eqref{hel2} from which one may derive the following
dispersion relation,
\begin{align}
&s^3 + k^2(\tfrac{7}{3}\alpha+\alpha_b)\,s^2
+\left[1+k^2+k^4\alpha(\tfrac{4}{3}\alpha+\alpha_b)\right]s \notag\\
 &\hskip5.7cm  +\alpha\,k^2\left[ 3(1+\beta) + k^2\right]=0. \label{dispersion}
\end{align}

In the long wavelength limit, $0<k\ll 1$, two instabilities may be
distinguished: the viscous instability and the viscous
overstability. The former possesses the growth
rate
$$ s= -3\alpha(1+\beta)\,k^2\,\, +\,\, \mathcal{O}(k^4),$$
and is unstable if $\beta<-1$. The instability will be
extinguished on short wavelengths and the critical $k$ upon which
marginal stability resides can be computed from
 $|k|= \sqrt{-3(1+\beta)}$. Wavelengths shorter than this value are viscously stable.
Table 1 states that a dense particulate ring possesses a positive
$\beta$ for all $\tau$, and so the viscous
mode will probably decay in real dense planetary rings.
 
 The
viscous overstability, on the other hand, possesses growth rates
\begin{align}\label{vogrowth}
  s= \pm i\,\left[1 +\tfrac{1}{2} k^2\right] 
 + \tfrac{1}{2}\left[
  3\alpha(1+\beta)-(\tfrac{7}{3}\alpha+\alpha_b)\right]k^2 \,\,+\,\,
  \mathcal{O}(|k|^3),  
\end{align}
and is hence oscillatory, corresponding to either standing waves or right or
  left-going travelling waves.
In the long wavelength regime, overstability emerges
  if the following criterion is satisfied:
\begin{align} \label{volin}
\beta>\frac{1}{3}\left( \frac{\alpha_b}{\alpha}-\frac{2}{3}\right),
\end{align}
(Schmit and Tscharnuter 1995). More generally, we have $\beta>\beta_c$, where $\beta_c$ summarises the
thermal properties of the system. 
A naive application of the condition \eqref{volin} to the data of Table 1 indicates
that dense planetary rings with optical depths of at least 1 are
viscously overstable (but compare with the detailed kinetic results of Latter and
Ogilvie 2008). Like the viscous instability sufficiently
short scales will be stabilised by pressure. The wavenumber of the marginal modes can be computed from the
quartic equation
\begin{align} \label{critk}
 \alpha(\alpha_b+\fourthirds\alpha)(\alpha_b+
  \seventhirds\alpha)k^4+(\alpha_b+\fourthirds\alpha)k^2    +(\alpha_b-\tfrac{2}{3}\alpha-3\alpha\beta)
=0. 
\end{align}

\section{Exact nonlinear solutions: uniform wavetrains}

\subsection{Introduction}

The linear theory is easy to establish; a more challenging task
 is to determine how the system evolves once it enters the nonlinear regime.
 At first,
small perturbations will grow independently and
exponentially in the form of overstable modes, but when
their amplitudes become sufficiently large they will interact. At this point
 the
evolution of the system becomes difficult to track and, as
with many hydrodynamical problems, we may be
 obliged to employ numerical simulations to describe its temporal and
 spatial variations.

Only one fully nonlinear hydrodynamical study
 has so far been undertaken, that of Schmit and Tscharnuter (1999). They
numerically evolved an isothermal, one-dimensional annulus of gas unstable to the viscous
 overstability. The temporal extent of their simulations reached some 10,000
 orbits and the radial extent some 4,000$H$ (corresponding to about 40
 km). These showed that the evolution of an overstable
 ring is disordered, but nevertheless manifests a
 characteristic spatial scale and a longer `beating' pattern (though this `beating'
 is perhaps an artefact of the finite size of the box).
 In addition, an
 overstable system exhibits an upward cascade of power to larger
 scales. If self-gravity is included, this process halts once a certain
 lengthscale is reached (equal to about $24H$ if $\alpha=\alpha_b=0.2628$ and
 $\beta=1.26$).
 If self-gravity is omitted they report
 that the upward transfer of power does not cease: by 10,000 orbits
 the system has injected most of its power to a scale of some $100H$.

 An important feature of these simulations is their
 enforcement of reflecting boundaries. Though the domain is large,
  these boundary conditions will ultimately impede the formation
 of structures associated with translational symmetry, such as travelling wave
 trains (at least globally). This is an important point because the weakly nonlinear analysis of
 Schmidt and Salo (2003) predicts that an overstable disk may support steady
 nonlinear waves, as do $N$-body simulations (Salo and Schmidt 2007).

We take the view that the saturation of the overstability will indeed be
  irregular (as in Schmit and Tscharnuter 1999)
 but that these
  spatio-temporal variations will manifest upon a bed of coherent
 nonlinear wavetrain solutions (similar to those revealed by Schmidt and Salo 2003).
 The latter solutions we may regard as `fixed
 points' in the disk's state space, i.e.\ the simplest nontrivial invariant solutions of
 the dynamics. The system's state trajectories we expect to wander around
 these points and in so doing exhibit much of their coherent structure.
 Similar phenomena characterise a number of diverse
  physical systems such as reaction--diffusion systems, flame fronts, and ocean waves (Whitham 1974, Infeld and Rowlands 1990, Doelman et
al. 2009), as well as higher dimensional turbulent systems, such as  pipe
 Poiseuille flow and plane Couette flow (Gibson et al.~2008 and references therein).

In this paper we make a start on investigating this idea. 
Our first task is to formally establish the existence of the axisymmetric nonlinear
travelling wavetrains, the fixed points,
from the fully nonlinear equations
\eqref{sigmad}--\eqref{ud}. This is what we undertake in this section. Once
this is done we can probe their linear stability (Section 5) and then
speculate on their role in the nonlinear dynamics generally (Section 6).

\subsection{Governing equations}
The axisymmetric nonlinear perturbation equations are 
\begin{align} \label{cuntnl}
& \d_t\sigma+u_x'\d_x\sigma+ \sigma\d_xu_x' =0, \\
& \d_t u_x' + u_x'\d_x u_x'-2u_y'= -\frac{1}{\sigma}\,\d_x f, \\
& \d_t u_y' + u_x'\d_x u_y' +\frac{1}{2}u_x' = -\frac{1}{\sigma}\,\d_x h,  \label{cuntnl2}
\end{align}
with 
\begin{align}\label{cuntnl3a}
&f= \sigma- (\alpha_b+\tfrac{4}{3}\alpha)\,\sigma^{1+\beta}\,\d_x u_x',
\\
&h= -\alpha \sigma^{1+\beta}\,(-\tfrac{3}{2}+ \d_x u_y'). \label{cuntnl3}
\end{align}
We propose that there exist wavetrain solutions with frequency
$\omega$ and wavenumber $\mu$, and which move with
the phase speed $c_p\equiv \omega/\mu$.
The
waveprofiles are stationary in a frame moving with this speed, and so
we let the density and velocity fields depend on
the single
 phase variable: 
\begin{equation} \label{theta}
\theta= \mu\,x-\omega t. 
\end{equation}
 In addition, so as to ensure a uniform wavetrain, it is
assumed that the density and velocity are $2\pi$ periodic in
$\theta$. In summary, we have
\begin{equation}\label{conde}
 \bm{Z}\equiv(\sigma,u_x',u_y'), \qquad
\bm{Z}(x,t)=\bm{Z}_0(\theta), \qquad
\bm{Z}_0(0)=\bm{Z}_0(2\pi). 
\end{equation}
Henceforth we drop the subscript 0.
Substituting this ansatz into the continuity equation \eqref{cuntnl} supplies
a first integral:
\begin{equation} \label{1stint}
 \sigma(u_x'-c_p)= \text{cst}.
\end{equation}
The constant can be computed from angular momentum conservation: if
we integrate the $y$-momentum equation over one period in $\theta$ we
obtain
$$ \frac{1}{2}\int_0^{2\pi}\sigma\,u_x'\,d\theta=0.$$
Integration of \eqref{1stint} yields $\text{cst}=-c_p$, which recommends
a new dependent variable $u=c_p-u_x'$. The surface density can then
be neatly expressed:
\begin{equation} \label{sig}
\sigma= \frac{c_p}{u}.
\end{equation}
 Also we relabel
$v=u_y'$ for notational uniformity. 

In order to compute the waveprofiles, $u$ and $v$, we must solve the momentum
equation for the perturbations. This
can be expressed as four ordinary differential equations:
\begin{align} \label{odes1}
\frac{d f}{d\theta} &= -c_p \frac{d u}{d\theta} + \frac{2}{\mu}\,\sigma\,v, \\
\frac{d h}{d\theta} &= c_p\,\frac{d v}{d\theta} -
\frac{c_p}{2\mu}\,\left(\sigma-1\right), \label{odes2}\\
\frac{d u}{d\theta} &=\frac{f-\sigma}{\mu(\alpha_b+\tfrac{4}{3}\alpha)\sigma^{1+\beta}},
\\
\frac{ d  v}{d\theta} &= \frac{1}{\mu}\left( \frac{3}{2} -
  \frac{h}{\alpha\sigma^{1+\beta}}\right). \label{odes3}
\end{align}
 This set is subject to the periodic boundary conditions presented in \eqref{conde}.

 The system \eqref{sig}---\eqref{odes3} is a
nonlinear eigenvalue problem with input parameters
$\alpha$, $\alpha_b$, $\beta$, and $\mu$ and with eigenvalue
$\omega$. Its solution can be accomplished numerically.
 These solutions we
present in Section 4.5, but first we establish some integral relations and some
 asymptotic results
which elucidate more clearly the various processes and balances at
work.

\subsection{Integral relations}

Two integral identities aid in the understanding of the solutions. The
first establishes the energy balance of the steady waves and shows that the important
terms are those responsible for viscous dissipation and viscous overstability. The second provides
an expression for the angular momentum transported by the waves.

We
derive two energy equations for the perturbations.
 The first governs the combined radial kinetic
energy and thermal energy:
\begin{align}
&\d_{t}\left(\tfrac{1}{2}\sigma (u_x')^2  + \sigma\ln\sigma
\right) \notag \\ &\hskip2cm  + \d_x\left( u_x'\left[ \tfrac{1}{2}\sigma (u_x')^2 
    +\sigma(\ln\sigma+1)-\Pi_{xx} \right]\right) \notag \\ &\hskip6cm  =2\sigma u_x' u_y' - (\d_x u_x')\,\Pi_{xx},\label{xenergy}
\end{align}
where $\Pi_{ij}$ is the viscous stress.
The second governs the azimuthal kinetic energy:
\begin{equation}\label{yenergy}
\d_{t}\left(\tfrac{1}{2}\sigma (u_y')^2
\right)+\d_x\left(\tfrac{1}{2}\,u_x'\,\sigma\,(u_y')^2 - u_y' \Pi_{xy} \right)=
-\tfrac{1}{2}\sigma\,u_x' u_y'- (\d_x u_y')\,\Pi_{xy}.
\end{equation}
 We now assume a steady wavetrain
solution with wavelength $\lambda$ and integrate over one wavelength. The
time derivatives and flux terms on the left sides of
Eqs~\eqref{xenergy}-\eqref{yenergy} vanish and we are left with the integrated
source terms on the right sides:
\begin{align*}
\int_0^\lambda [2\sigma u_x' u_y' - (\d_x u_x')\Pi_{xx} ] \, dx =0, \qquad
\int_0^\lambda [\tfrac{1}{2}\sigma u_x' u_y' + (\d_x u_y') \Pi_{xy}]\, dx=0.
\end{align*}
These are combined so as to remove the `Reynolds stress' term, and this
furnishes us with the key balance:
\begin{align} \label{visccont2}
\int_0^{\lambda}\sigma \left\{ \,6\nu \d_x u_y' \,-\, \left[
    (\nu_b+\tfrac{4}{3}\nu)\,(\d_x u_x')^2 +4\nu (\d_x u_y')^2 \right] \,\right\}\,dx\,=\,0.
\end{align}
In order to satisfy this equation, the first term, associated with viscous
 overstability,
 must integrate to a positive value and balance the second viscous dissipation
 term which is always negative. The equation is an energy
 constraint that a uniform amplitude wavetrain must satisfy:
 kinetic energy removed by viscous dissipation must be exactly
 balanced by kinetic energy injected from linear viscous overstability.
As such, Eq.~\eqref{visccont2} provides a means by which we can calculate the
 rate of strain (and nonlinear amplitude) of a wavetrain. 

An expression for the total angular momentum flux over a
wavelength is easy to derive. We find the flux is equal to
\begin{align} \label{angm}
 \frac{1}{\lambda}\int_0^{\lambda} (\sigma u_x' u_y' -\Pi_{xy})\,dx
                       = \frac{1}{2\pi}\int_0^{2\pi} (h- c_p\, v)\, d\theta.
\end{align} 
where we have used \eqref{odes1}. In order to isolate the contribution from
the nonlinear wave we must deduct the equilibrium angular momentum flux
$\frac{3}{2}\alpha$, which just issues from the combination of Keplerian shear
and viscosity. The wave flux, we find, always points outwards and
 never exceeds that of the background viscous stress. The nonlinear waves 
 never transport angular momentum inward unlike
 certain vigorously forced density waves (Borderies, Goldreich and
 Tremaine 1983).

\subsection{Asymptotic description in the limit of long wavelength}

Consider Eqs \eqref{sig}-\eqref{odes3} and 
suppose that their solution possesses a particularly long
wavelength, meaning $0<\mu\ll 1$. In addition, let the waves oscillate
near the epicyclic frequency so that
$\omega=1+\mathcal{O}(\mu^2)$. Because we must have
$\sigma=\mathcal{O}(1)$, it follows from Eqs \eqref{sig} and \eqref{odes1}
that $u$ and $v$ are $\mathcal{O}(\mu^{-1})$ and that
the collective effects of pressure and
viscosity (and self-gravity) are subdominant in the force
balances. Consequently, the wave profile to leading order is
determined by rotation and shear, a simplification that permits the momentum equation to
 be solved analytically:
\begin{align} \label{as1}
u_x' \approx -\mu^{-1}\,q\, \cos\vartheta, \qquad u_y' \approx -\tfrac{1}{2}\mu^{-1}\,q\,\sin\vartheta
\end{align} 
where the new phase variable $\vartheta(\theta)$ proceeds from the transcendental equation
\begin{equation} \label{as2}
\theta-\theta_0= \vartheta+q\sin\vartheta.
\end{equation}
Here $q$ and $\theta_0$ are constants of the integration.
The scaled amplitude $q$ must be positive and less than 1 (or else
neighbouring streamlines cross) and quantifies the rate of strain associated with the
nonlinear wave, and hence the quantity of energy dissipated.
It can be determined from Eq.~\eqref{visccont2}
or at higher order in the asymptotic expansion. 

Note that the solution \eqref{as1} is self-similar with respect to
wavelength; it follows that our disk model can support nonlinear wavetrains of
arbitrary amplitude and wavelength, at least formally.
In the limit of large
amplitude, the waves carry only a finite
angular momentum flux, as is clear when
the
waveprofiles of \eqref{as1} are substituted into \eqref{angm}.
In particular, the coefficient of the
$\mu^{-1}$ term, which is associated with the Reynolds stress, integrates to zero.

The next order in the expansion appears intractable. But, if we
shift to a pseudo-Lagrangian description we can simultaneously generalise and
simplify the problem.  
At leading order
collective effects are small and the
fluid elements almost exactly trace out epicycles, which may then be removed
by averaging.
 We fully work out the
pseudo-Lagrangian theory in Appendix A and summarise the main results here: the
nonlinear dispersion relation and the equation for $q$.
The nonlinear dispersion relation is
\begin{equation} \label{nldispt}
\omega=1  +\mu^2 F_1(q)+\mathcal{O}(\mu^3),
\end{equation}
in which pressure enters through the $F_1$ function
\begin{equation} \label{HHt}
F_1(q)= q^{-2}[(1-q^2)^{-1/2}-1].
\end{equation}
Steady waves require $\omega$ to be real in
 in Eq.~\eqref{asomega}, which returns
  a transcendental equation for $q$. One obtains
\begin{align}
(\alpha_b+\tfrac{16}{3}\alpha)\frac{1}{\widetilde{q}}\,\text{P}^{(1)}_\beta(\widetilde{q})
- 3 \alpha\beta\,\text{P}^{(1)}_{1+\beta}(\widetilde{q}) + 4\alpha\,q\,\text{P}^{(2)}_{1+\beta}(\widetilde{q})=0 
 \label{FFt}
\end{align}
with $\widetilde{q}=(1-q^2)^{-1/2}$ and
where $\text{P}^{(m)}_{\eta}$ is the associated Legendre function of the first kind and
type 3 (see Appendix A).
 For non-integer $\beta$, Equation \eqref{FFt} must be solved numerically.

\subsection{Numerical computation of wavetrain solutions}

In this section solutions to \eqref{sig}---\eqref{odes3} are computed
 numerically two ways: by the shooting and the pseudospectral methods.
  Our shooting algorithm
integrates the equations in $\theta$ over one period with a 4th/5th order Runge-Kutta scheme
with adaptive stepsizing.  The algorithm converges onto the correct
solution
 via the application of the
four-dimensional Newton's method,
 which enforces the four boundary conditions at $\theta=2\pi$. 
On the other hand, the pseudospectral method partitions
 the $\theta$
 domain into $N$ points and evaluates the functions' spatial derivatives
  via excursions into Fourier space. Consequently, the governing equations
 boil down to $4N$ nonlinear algebraic equations for the 4 functions
 evaluated at the $N$ points, plus the eigenvalue $\omega$. Because of translational
 symmetry we may specify $u=0$ at $\theta=0$ which means we have
 $4N$ equations and $4N$ unknowns. The equations are solved using the
 multidimensional Newton's method.

The input parameters are $\alpha$,
$\alpha_b$, and $\beta$ which arise from the viscosity prescription
\eqref{nupresc}, and $\mu$, the wavenumber of the wavetrain we hope to
compute. We present results for one set of parameters only, those corresponding to
an optical depth $1.5$ (cf.\ Table 1), as there is little
variation in the qualitative features as we vary the viscous parameters.
 Therefore, $\alpha=0.342$, $\alpha_b=0.681$, and $\beta=1.19$.

For these fixed parameters we examined different wavenumbers $\mu$. We found
travelling wave solutions for all $\mu<\mu_c$, where $\mu_c$ corresponds to
the marginal $k$ of linear viscous overstability in
Eq.~\eqref{critk}. For the above parameters $\mu_c=0.7295$.
If the disk was not viscously overstable (for instance, if we took the parameters
associated with $\tau=0.5$) then no nonlinear wavetrains could be found.
 This, of course, confirms that the 
wavetrains rely on the viscous overstability to
sustain them against conventional viscous loss. For $\alpha_b$ and $\alpha$
 constant, the branch of solutions is
 supercritical with respect to $\beta$.

\begin{figure}[!ht]
\begin{center}
\scalebox{.55}{\includegraphics{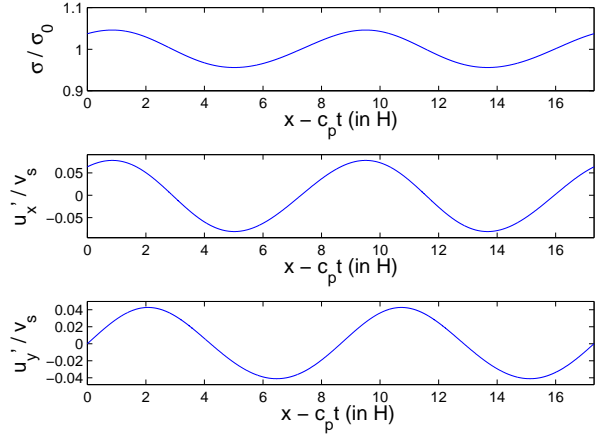}}
\caption{\footnotesize{Two periods of a small amplitude nonlinear wavetrain
    as calculated by the numerical eigenvalue problem, \eqref{sig}-\eqref{odes3}. 
 The waveprofile is plotted in the spatial coordinate of a
    frame comoving with the wave.
The parameters correspond to $\tau=1.5$ from Table 1:
    $\alpha=0.342$, $\alpha_b=0.681$, and $\beta=1.19$. The wavenumber chosen is
    $\mu=0.726$ ($\lambda=8.65$) which is just below the critical $\mu_c$. The
 wavetrain profile coincides with the marginally stable overstable \emph{linear} mode of the
 homogeneous state of Keplerian shear.}}
\end{center}
\end{figure}

\begin{figure}[!ht]
\begin{center}
\scalebox{.55}{\includegraphics{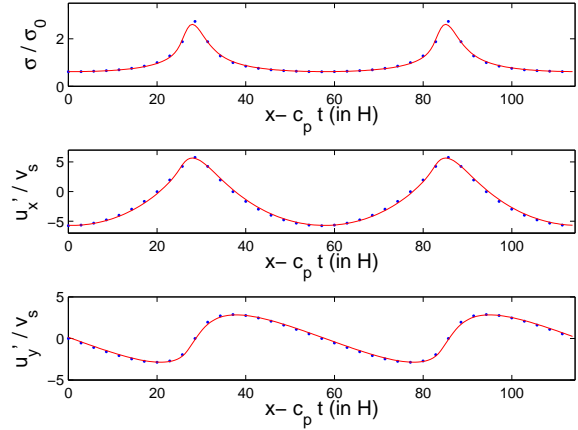}}
\caption{\footnotesize{Two periods of a large amplitude wavetrain; both asymptotic and numerical
    solution are plotted.
 The solid curve is the numerical solution to \eqref{odes1}--\eqref{odes3}, the points
    represents the asymptotic solution \eqref{as1}--\eqref{FFt}. Parameters
    correspond to $\tau=1.5$, as in Fig.~1, but
    now with $\mu=0.11$, so that $\lambda=57.1$.}}
\end{center}
\end{figure}

\begin{figure}
\begin{center}
\scalebox{.5}{\includegraphics{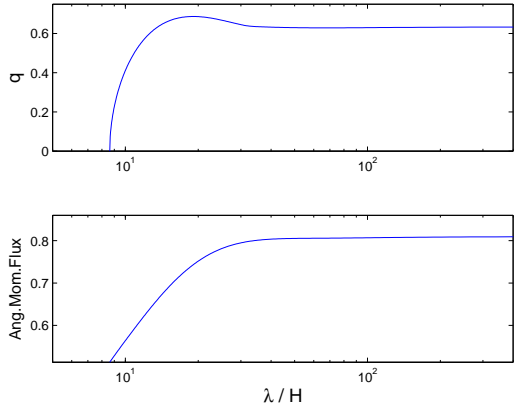}}
\caption{\footnotesize{Panel (a) displays the scaled velocity amplitude of the
    nonlinear waves
    $q$ as a function of $\lambda=2\pi/\mu$.
   Near the critical wavelength, $q$ is almost
    zero and we are in the linear regime. From there it increases steeply
    and approaches a constant value, which agrees with the
    asymptotic $q$ computed from \eqref{FFt}, $q_0=0.63339$. Panel (b)
    displays the total angular momentum flux.  At amplitudes near criticality, the waves carry
    negligible amounts of angular momentum and the background shear flow
    dominates the transport. In contrast, larger amplitude wavetrains can carry
    appreciable amounts of up to a third of the background $3\alpha/2$. Recall
    $\alpha=0.342$.}}
\end{center}
\end{figure}

\begin{figure}
\begin{center}
\scalebox{.5}{\includegraphics{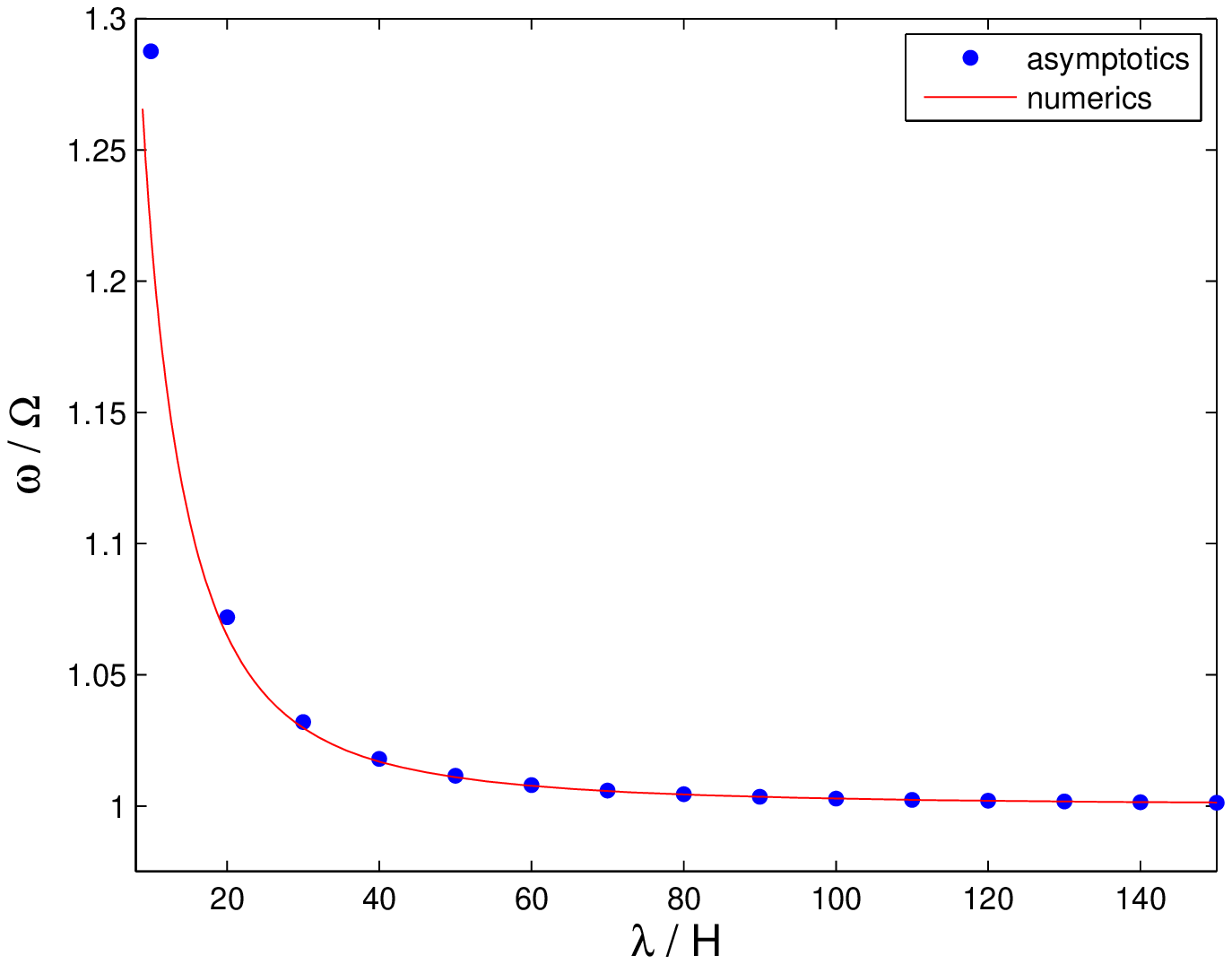}}
\caption{\footnotesize{The nonlinear wavetrain frequency
    $\omega$ is graphed as a function of $\lambda$. The solid
    curves represent the numerical solution and the points the prediction of
    the asymptotic theory, \eqref{nldispt}.}}
\end{center}
\end{figure}

Wavetrains
with $\mu$ near $\mu_c$ possess small amplitudes, as is shown
in Fig.~1. In this small amplitude
`linear regime' the solutions we compute connect, naturally, to the marginal linear
overstable modes, which are necessarily sinusoidal.

As $\mu$ decreases ($\lambda$ increases) the nonlinear amplitudes increase.
 For sufficiently
long wavelengths the self-similar asymptotic profiles of \eqref{as1}---\eqref{FFt}
 provide a good approximation
to the solutions, as we can see in Fig.~2.  The profiles of
these long waves resemble those of spiral density waves, as computed
by Shu et al.~(1985a, 1985b) and Borderies et al.~(1983, 1986), and also
axisymmetric density waves in accretion disks (Fromang and Papaloizou
2007). The characteristic `cusp'-like profile seems a recurrent feature of
 density waves dominated by the inertial forces.

In addition to these typical profiles we show how the scaled velocity amplitude $q$ varies with
wavelength $\lambda=2\pi/\mu$ in Fig.~3a. Recall that the actual velocity amplitude is
equal to $q/\mu$ (cf.\ Eq.~\eqref{as1}).
By about $\lambda\approx 30\,H$ the amplitude $q$ begins to
approach the value computed by the asymptotics, Eq.~\eqref{FFt}.
 For the parameters in which we are interested, the asymptotic result becomes
an acceptable approximation for wavelengths above $20-30\,H$. 

In Fig.~3b we plot the angular momentum flux density over one period in
the wavetrain as a function of $\lambda$. We use the expression \eqref{angm}. 
The small amplitude waves transfer negligible angular momentum, so
when $\mu\approx \mu_c$ the angular momentum flux is approximately
that of the unperturbed disk, specifically $(3/2)\alpha$. Longer
wavelength, larger amplitude wavetrains, however, can carry appreciable
quantities above that of the background shear, but they may never exceed
it. The angular momentum flux asymptotes to a constant
value for large wavelengths, in agreement with the asymptotic theory.

Lastly, in Fig.~4 the eigenvalue $\omega$ is plotted against $\lambda$ for the same
set of parameters.
 The longer the wavelength the closer the
frequency is to the epicyclic with pressure ensuring the oscillations are a
little faster.
These behaviours are predicted by the 
asymptotic nonlinear dispersion relation which appears to provide an adequate
approximation by about $\lambda\approx 30-40\,H$.

\subsection{Discussion of neglected physics}

The wave solutions have been computed by a relatively simple fluid model, but
we feel that their existence and gross features are not captive to its
 assumptions. The asymptotic analysis shows that self-gravity (a
collective effect) will not alter the leading order wave profiles.
 Self-gravity will, however, alter the nonlinear dispersion
relation \eqref{nldispt}, adding a term proportional to $|\mu|$. Also, by
exacerbating the linear viscous overstability it will `shift' the curve in
Fig.~3a to the left, which means that for a given $\lambda$ a wavetrain will be
`more nonlinear' (its amplitude will be greater) when self-gravity is present than when
it is not. On the other hand, neglected thermal effects help stabilise the
linear instability (Spahn et al.~2000) and so the curve in Fig.~3a may shift back to the right
again. Thermal effects will also alter the pressure contribution to the
nonlinear dispersion relation embodied by the function $F_1$.

It is more difficult to ascertain the effect of vertical structure and 
vertical motion on the
solutions. The linear modes may be stabilised on
intermediate scales due to their development of vertical shear (see Latter and
Ogilvie 2006b), thus shifting Fig.~3a a little to the right. But because planetary
rings are so very thin this is probably not important.
 There will be additional effects tied to the vertical motion, the
 `breathing', which allied with excluded volume effects at wave peaks, could lead to the
 vertical ejection of particles and the relaxation of the pressure and stress
 in real planetary rings (`splashing'). But how this plays out on moving wavecrests
 is difficult to forecast.
The role of a
 nonlinear stress/strain relationship is also difficult to judge, though at
 the very least it
 will probably lead to greater effective $\beta$s and hence shift the $q$
 curve to the right.

In summary, we have demonstrated the existence and elaborated on the
properties of a family of exact nonlinear solutions to the governing
equations \eqref{cuntnl}-\eqref{cuntnl2}.
 The solutions take the form of
steady, travelling density waves propagating in either radial 
direction. They may also be regarded as fixed points or periodic orbits
in the state space of the system.
The members of this family can be distinguished by their wavenumber $\mu$ (or wavelength
$\lambda$) upon which only one restriction holds, $\mu<\mu_c$, where $\mu_c$
is the critical wavenumber of linear viscous overstability. As such, an infinite
number of solutions formally exist on arbitrarily long lengthscales; however
above a certain lengthscale the model assumptions will certainly break down.

\section{Linear stability of the nonlinear wavetrains}

The question now is: what role do the nonlinear solutions
play in the evolution of an overstable disk? 
To make a start on this we need
to determine the
\emph{linear stability} of the nonlinear solutions themselves. If the system is to
settle on a wavetrain then it must be at least linearly stable. On
the other hand, if it is linearly unstable then we are assured that
the system will not settle there (though this does not mean it plays no role
in the dynamics). We find, in fact, that there exists
 a critical wavelength $\lambda_\text{st}$ above which linear stability is assured. Wavetrain solutions with
 $\lambda<\lambda_\text{st}$ are unstable and solutions with
 $\lambda>\lambda_\text{st}$ are stable.

These results are established by perturbing the wavetrain
 solution by a small disturbance and subsequently
 solving the eigenvalue problem that results for its growth
 rate.
 The procedure requires the solution of a Floquet boundary value
 problem which we attack analytically in the asymptotic limit
 of $0<\mu \ll 1$ (Appendix A) and numerically for general $\mu$.

\subsection{The linear eigenvalue problem}

 Consider the wavetrain associated with $\mu$ 
and parameters $\alpha$, $\alpha_b$, and $\beta$. 
We denote this basic state by $\bm{Z}_0(\theta)=(\sigma_0, u_{x0},
u_{y0})(\theta)$, with the dependence on $\mu$ assumed.
 On this solution we superimpose a small disturbance so that
$$ \bm{Z}= \bm{Z}_0(\theta) + \bm{Z}_1(t,\theta), $$
where $\|\bm{Z}_1\| \ll 1$. Upon substituting this ansatz into the
governing equations \eqref{sigmad}-\eqref{ud}, employing the definition
\eqref{theta}, and linearising, the system returns the boundary value problem:
\begin{equation} \label{lin1}
\d_t \bm{Z}_1 = \mathcal{L}_0(\theta)\,\bm{Z}_1
\end{equation}
where $\mathcal{L}_0$ is a second-order differential operator in $\theta$ with
$2\pi$-periodic coefficients.

Next we assume $\bm{Z}_1(t,\theta)=e^{st}\,\widetilde{\bm{Z}}_1(\theta)$ which reduces
\eqref{lin1} to a Floquet eigenvalue problem for the growth rate $s$, owing to
the periodicity of
$\mathcal{L}_0$. In order to compute
the spectrum we consequently make the Floquet  ansatz,
\begin{equation}
\widetilde{\bm{Z}}_1 = e^{i(k/\mu)\theta}\, \hat{\bm{Z}}_1(\theta),
\end{equation}
where $\hat{\bm{Z}}_1$ is a $2\pi$ periodic function of $\theta$, the
Floquet exponent is $(k/\mu)$, and $k$ is the (real) wavenumber of the envelope (both $s$ and
$k$ are different to those
appearing in Section 3). Once substituted back into \eqref{lin1}, one can
numerically solve the ensuing eigenproblem for $\hat{\bm{Z}}_1$ and the
eigenvalue $s$ if the parameter $k$ is stipulated.

We now present the full set of linear equations governing the initial evolution
of the small disturbances. Instead of rendering the problem in the form of Eq.~\eqref{lin1}, 
we express it as five simpler first-order differential equations for
 $\hat{\sigma}_1$, $\hat{u}_1= c_p- \hat{u}_{x1}$,
$\hat{v}_1=\hat{u}_{y1}$, $\hat{f}_1$,
and $\hat{h}_1$. This form facilitates the numerical calculation. The hats
will now be dropped.
 The five ODEs are
\begin{align} \label{lineq1}
\frac{d\sigma_1}{d\theta} &= \frac{1}{u_0}\left[ \left( \frac{s-ik u_0}{\mu}-\frac{d u_0}{d\theta}
  \right)\sigma_1 - \left(\frac{d\sigma_0}{d\theta}+\frac{ik\sigma_0 }{\mu} \right)u_1-
  \sigma_0\,\frac{d u_1}{d_\theta} \right]\,, \\
\frac{d u_1}{d\theta} &= -\frac{ik}{\mu}\,u_1 -(1+\beta)\,\frac{d u_0}{d\theta}\, \frac{\sigma_1}{\sigma_0}
- \frac{\sigma_1-f_1}{\mu(\alpha_b+\tfrac{4}{3}\alpha)\sigma_0^{1+\beta}}\,,
\\
\frac{d v_1}{d\theta} &= -\frac{ik}{\mu}\,v_1 -(1+\beta)\,\frac{d v_0}{d\theta}\,
\frac{\sigma_1}{\sigma_0}+\frac{\tfrac{3}{2}\alpha(1+\beta)\sigma_0^{\beta}\,\sigma_1-h_1}{\mu\alpha
  \sigma_0^{1+\beta}}\,, \\
\frac{d f_1}{d\theta} &= -\frac{ik}{\mu}\,f_1 +\sigma_0  \left[ \left(\frac{s-ik
  u_0}{\mu} - \frac{d u_0}{d\theta} \right)\,u_1 
 - u_0\,\frac{d u_1}{d\theta} + \frac{2}{\mu}\,v_1 + \frac{d
  f_0}{d\theta}\,\frac{\sigma_1}{\sigma_0^2} \right]\,, \\
\frac{ d h_1}{d\theta} &= -\frac{ik}{\mu}\,h_1- \sigma_0\left[ \frac{s- ik u_0}{\mu}\,v_1- u_0\,\frac{ d v_1}{d\theta}
  - \left(\frac{d v_0}{d\theta}+ \frac{1}{2\mu}\right)\,u_1
  -\frac{d h_0}{d\theta}\,\frac{\sigma_1}{\sigma_0^2} \right], \label{lineq5}
\end{align}
and these hold on the domain $\theta\in [0,2\pi]$ 
subject to the periodic boundary conditions, $\sigma_1(0)=\sigma_1(2\pi)$,
$u_1(0)=u_1(2\pi)$, etc. Before we undertake the calculation, 
we need to specify the material parameters $\alpha$, $\alpha_b$,
$\beta$, the wavenumber of the underlying wavetrain $\mu$ (and hence the basic
state $\bm{Z}_0$),
 and lastly the wavenumber of the
disturbance $k$.

\subsection{Numerical results}

The equations for the equilibrium \eqref{sig}-\eqref{odes3} and the linear stability
\eqref{lineq1}-\eqref{lineq5}
are solved simultaneously using the shooting method.
We restrict the parameter values to those corresponding to optical depths of
1-2
 and similar parameters. For these values we witness little
qualitative change in the results. For more remote values, especially $\alpha_b/\alpha$
small or zero, we find different stability behaviour but this will not
be presented here. 

Once $\alpha$, $\alpha_b$, and $\beta$ are fixed, we investigate
 how the various linear modes depend on the wavenumber of the underlying
wavetrain $\mu$, as a function of
$k$. The reader should take care to distinguish between $\mu=2\pi/\lambda$, the
 wavenumber of the perturbed equilibrium, and $k$, the
envelope wavenumber of the perturbation. 
Normally, one need only let $k$ vary between $0$ and $\mu/2$; values of $k$
outside this range merely reproduce modes already
computed. However, in this section we let $k$ take values between $0$ and
$\mu$. This choice lets us display and interpret the results in a clearer (and
more familiar) way, while permitting us to connect the results directly with
the homogeneous disk case.

The numerical solution establishes the existence of 
\begin{itemize}
\item a `generalised'
viscous instability mode, which always decays at a rate proportional to
$\alpha k^2$;
\item  two
 `generalised' viscous overstability modes which can either grow or decay at
 a rate proportional to $\alpha k^2$ and which oscillate near the epicyclic
 frequency;
\item  modes which decay rapidly, at a rate of order $\alpha$ and which
also oscillate at the epicyclic frequency.
\end{itemize}
The two modulational
 overstable modes appear as right-going and left-going waves. They 
generally possess different growth rates because
the equilibrium background (a nonlinear wavetrain propagating either left or
 right)
 no
longer supports the left/right symmetry exhibited by the homogeneous disk. We
are primarily interested in these modes, as they are the only ones which
are capable of growing.  We refer to them
as `modulational' because they exist as long modulations of the phase and amplitude
of the equilibrium wavetrains. Thus the fastest growing overstable modes possess
wavelengths significantly longer than the wavelength of the equilibrium
wavetrains.

\begin{figure}[!ht]
\begin{center}
\scalebox{.6}{\includegraphics{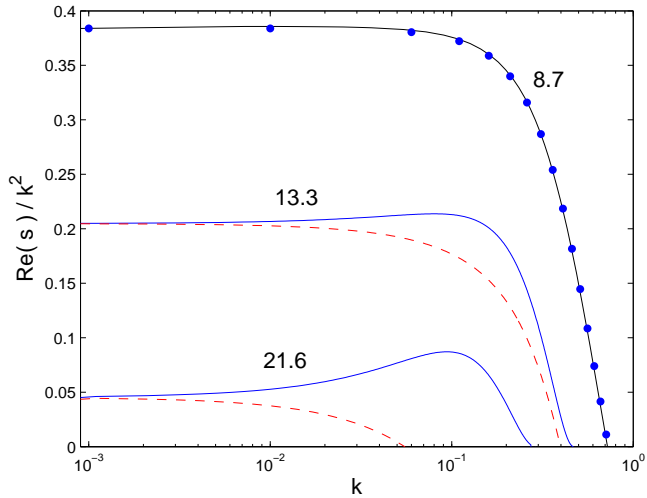}}
\caption{\footnotesize{ Real part of the scaled growth rate $s/k^2$ as a
    function of the wavenumber $k$ for modulational overstable modes. The
    solid lines correspond to the modes propagating against the underlying
    wavetrain; dashed lines correspond to modes propagating with.
    Three background
    equilibria are selected: nonlinear wavetrains with wavelengths $8.7$, $13.3$, $21.6$.
   The material parameters of the disk are taken from Table 1 for
    $\tau=1.5$. In addition, we plot the stability curve for the homogeneous equilibrium
    (from Eq.~\eqref{dispersion}) as points. The amplitude of the $\lambda=8.7$ nonlinear wavetrain
    is very small (being close to the critical wavelength) and thus the disk
    is in the linear regime of the homogeneous disk. Therefore the stability curve
    in these two cases agree. Larger $\lambda$ and wavetrain amplitudes impede modulational
    overstability. But modes propagating against the background wavetrain are
    more unstable than those propagating in the same direction. In the
    homogeneous disk the growth rates are the same.}}
\end{center}
\end{figure}

First, in order to check the numerical results, we examined the stability of
wavetrains of very small amplitude, i.e.\ those which possess a wavenumber $\mu$ near
the critical value $\mu_c$. Recall that $\mu_c$
is the wavenumber of marginal viscous overstability (see Eq.~\eqref{critk}) and
simultaneously of the very existence of nonlinear wavetrain solutions (see
Fig.~3).
 In this limit the
nonlinear wavetrain solution, being of small amplitude, can be simply absorbed into
the 
perturbation field and the equilibrium treated as the homogeneous state.
In this limit the growth rates for both the `generalised' viscous
instability and `generalised' viscous overstability coincide exactly with their
homogeneous disk counterparts (Eq.~\eqref{dispersion}),
 which provides the check on the
numerics. The good agreement obtained between the two approaches is evident in
Fig.~5 which plots the numerical growth rate of the modulational overstability, for the
small amplitude wavetrain $\lambda=8.7$, alongside the analytic growth rate
obtained from the homogeneous problem \eqref{dispersion}.

\begin{figure}[!ht]
\begin{center}
\scalebox{.6}{\includegraphics{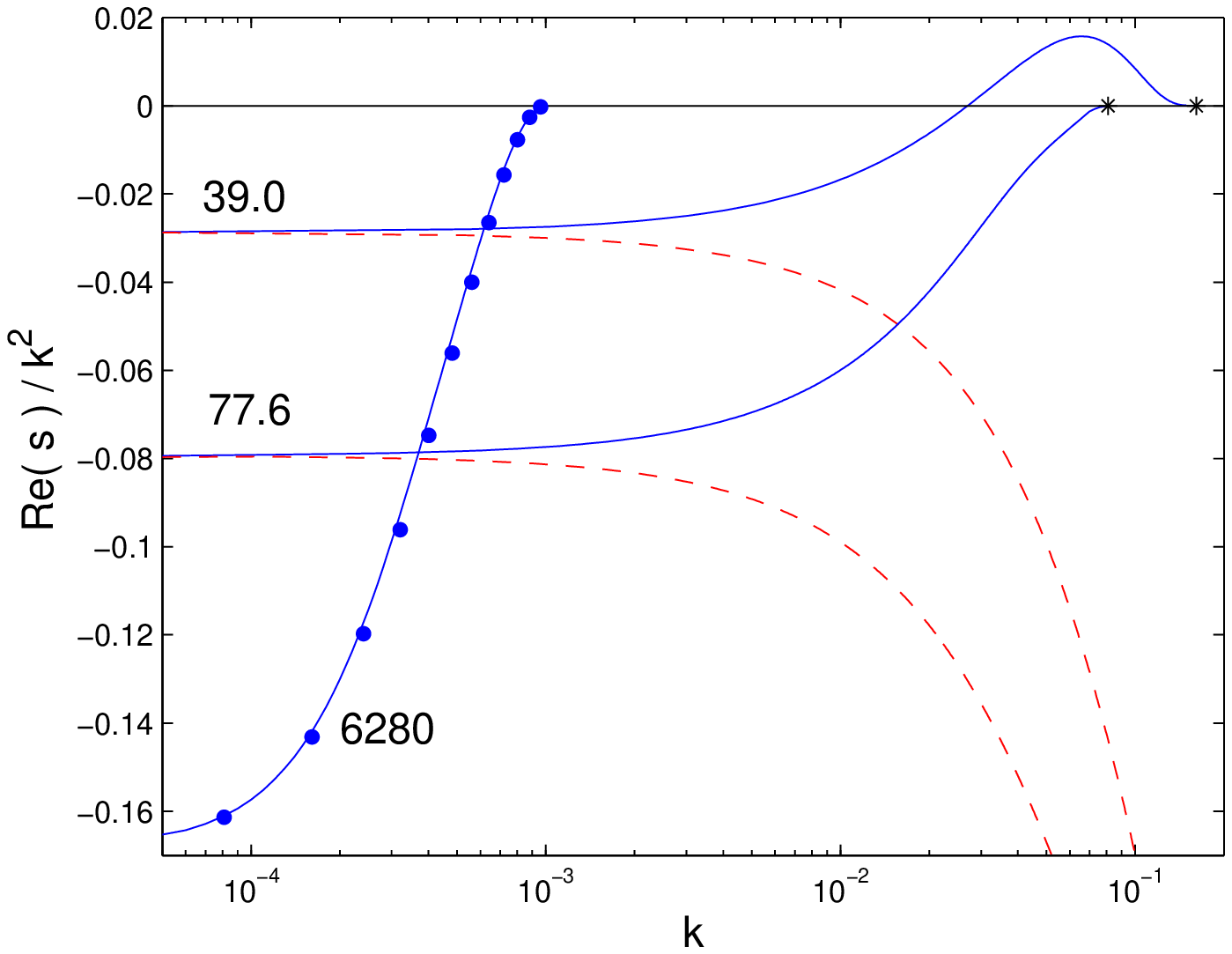}}
\caption{\footnotesize{Continuation of Fig.~5. Real part of the scaled growth rate $s/k^2$ as a
    function of $k$ for the modulational
    overstable modes. We present results for three background wavetrains: $\lambda=39.0$,
    $\lambda=77.6$, and the very long $\lambda=6280$ case.
 The material parameters and notation are the same as
    the previous figure and the range of $k$ is $0$ to $\mu$.
Stars indicate the point $k=\mu$ in both cases.
   The $\lambda=39.0$ graphs show the persistence of instability on
    intermediate $k$ for the counter-propagating mode, and stability for its oppositely
    moving twin. By $\lambda=77.6$ both modes have stabilised. In
    addition, we plot the prediction of the large $\lambda$ asymptotic theory
    (see Appendix A) as points,
    and find good agreement between it and the $\lambda=6280$ case.}}
\end{center}
\end{figure}

Next we gradually increase the wavelength of the background wave $\lambda$
and check its stability at each step. Some of these
results we present in Figs 5 and 6. These figures show
the growth rates of the two modulational overstability modes as a function of $k$ for
several representative wavetrains. The model parameters correspond to
$\tau=1.5$ in Table 1. In the first figure, the stability of $\lambda=8.7$,
$13.3$ and $21.6$ wavetrains are plotted, with the growth rate of the
overstability mode propagating \emph{with} the underlying wavetrain
represented by the dashed line, and the growth rate of the mode propagating
\emph{against} by the solid line. As is plain, increasing the $\lambda$ of the
background (a) impedes both modes' growth and (b) breaks the left/right
symmetry of the problem so that the dispersion
relations of the two modes no longer coincide. In particular, growth of the countermoving
 mode is exacerbated
 on an intermediate range of $k$. In the limit of very small $k$ the growth
 rates of the two modes are the same.

 From Fig.~6 we see that both modes stabilise when the underlying wavetrain is
 sufficiently long. Though the critical $\lambda$ at which this happens 
 is different for the two modes (because of the exacerbated growth of the countermoving
 mode at
 intermediate $k$). For the parameters chosen,
 the modulational overstable
 mode travelling with the background stabilises when $\lambda\approx
 27.0$ (or 119 m, using the dimensions of Table 1), while the
 countermoving mode stabilises when $\lambda\approx 52.8$ (or 234 m).
 We conclude that the nonlinear wavetrains'
 stability criterion is controlled by the countermoving overstability
 mode. The critical wavelength above which a wavetrain is stable we denote by
 $\lambda_\text{st}$. And so sufficiently long wavetrains are
 linearly stable to all disturbances, which is in agreement with
 the asymptotic theory of Appendix A.

Lastly, the stability of very long wavetrains was checked so as to
compare with the asymptotic linear stability and provide a second check
on the numerics. In Sections A.4.1 a dispersion relation was computed
giving the growth rate of the dominant modulational overstability mode. It
is plotted with points in Fig.~6 against the prediction of the numerical eigensolution
 for a wavetrain with $\mu=0.001$. Though the agreement is good in this
 extreme case, the asymptotics provide an adequate approximation only for
very long $\lambda$.

\subsection{Summary}

The linear theory is rather subtle and may repay extended study.
For now, the essential point to take away is that there exist both linearly stable and
linearly unstable wavetrain solutions. Wavetrains possessing a
sufficiently long wavelength, namely $\lambda>\lambda_\text{st}$, are stable, while
those that are shorter are unstable. The mode which governs the stability
properties is the `modulational viscous overstability', in particular, the mode
which travels in the opposite direction to the underlying wavetrain. 
Those wavetrains that are unstable can be characterised
as `saddle points' in the phase space because they possess the unstable eigenmodes (the
modulational overstabilities, which
grow slowly, no faster than $\alpha k^2$), and a set of stable eigenmodes (some of which
decay speedily, as $\alpha$).

This general behaviour is repeated for all realistic parameters we tried.
If we turn to the parameters listed in Table 1, a ring with $\tau=1$ yields
$\lambda_\text{st}=50.2$ (165 m), a ring with $\tau=1.5$ yields $\lambda_\text{st}= 52.8$ (234
m), and a ring with $\tau=2$ yields $\lambda_\text{st}= 117$ (640 m). The much longer
value in the last case is due to the system's sensitivity to the $\beta$
parameter. For fixed $\alpha$ and $\alpha_b$ we find that $\lambda_\text{st}$
is a steeply increasing function of $\beta$, a trend that is summarised in
Table 3. The critical lengths at $\tau =1$-$1.5$
compare suggestively to the scales of quasi-periodic microstructure observed by
Cassini, which are some 150-220 m (Colwell et al.~2007, Thomson et al.~2007).
But it should be noted that important
physics remains missing and the stability estimates are subject to some revision.
In particular, the results for $\tau=2$, and greater $\beta$ generally, are probably
blemished by neglect of certain physical processes (self-gravity, for
example). Nevertheless, the overall consistency
is encouraging.

\begin{table}[!ht]
\begin{center}
\begin{footnotesize}
\begin{tabular}{|c|c|}
\hline 
 $\beta$ & $\lambda_\text{st}$  (in $H$) \\ 
\hline
1.1 & 41.9 \\
\hline
1.19 & 52.8 \\
\hline
1.3 & 71.0 \\
\hline
1.4 & 91.7 \\
\hline
1.5 & 121 \\
\hline
\end{tabular}
  \caption{\footnotesize{The critical $\lambda$ of stability for nonlinear wavetrains as a
 function of $\beta$. The other parameters $\alpha$ and $\alpha_b$ are held
 fixed: $\alpha=0.342$ and $\alpha_b=0.681$. The case $\beta=1.19$ corresponds
 to an optical depth of 1.5, larger values of $\beta$ correspond to more
 optically thick disks. For instance, the $\beta=1.5$ case is associated with
 $\tau=2$ (Schmidt et al.~2001).}}
 \end{footnotesize}
\end{center}
\end{table}

\section{General nonlinear dynamics}

Armed with the linear stability results, we return to the question posed
 earlier:
 what role does this family of
`fixed points' play in the general dynamics of an overstable disk?
The simplest outcome would be for an overstable disk to migrate from the
 homogeneous state to the shortest
 stable wavetrain solution (with $\lambda=\lambda_\text{st}$); but this is far from guaranteed.
 The dynamics is likely to be more complicated, with the system
 flirting with a number of stable nonlinear solutions and thus exhibiting irregular
 time and space dependent variations.
In this section we explore this behaviour with some schematic ideas and analogies.
 We will not bury ourselves in technical
details; this we leave for later, when we have a more complete set of
numerical results at hand.

First, we offer an explanation of the cascade of power to longer
lengths that has been observed in nonlinear
hydrodynamical and $N$-body simulations. The analysis is framed in the language of
dynamical systems theory and though `hand-wavey' should be
straightforward to check numerically. Next we examine the nonlinear
interactions between linearly stable wavetrains of different wavelength, by
computing slow variations in the wavetrains' phase. It can be
shown that the disk supports modulational `shock' and `source' structures, whereby the
wavenumber and/or amplitude of a wavetrain undergoes a discontinuity. Lastly,
we draw analogies between a viscously overstable disk and the dynamics of the
complex Ginzburg-Landau equation, and subsequently make a few additional predictions
about the likely nonlinear behaviours associated with overstable disks.

\subsection{The cascade to longer lengthscales}

In Fig.~7 we have drawn a schematic diagram of the state space of the overstable
disk. The state space is infinite-dimensional, and, needless to say, somewhat
difficult to visualise, but our crude two-dimensional projection
 raises a few interesting points and predictions. This
representation of the dynamics treats the time-evolution of the
 state vector $\bm{Z}$ as a trajectory, tracing the smooth transition
of the system from
 one state to another. Equilibria appear as
invariant fixed points: trajectories that begin there stay there, and their
linear stability decides if trajectories that pass infinitesimally close
fall into the fixed point or are repelled. The state of homogeneous
Keplerian shear is such a fixed point, and when it is overstable it behaves like
a saddle. 
Nearby trajectories
will be deflected away from the point along those directions associated with the
overstability modes. Simultaneously, the same trajectories will be drawn towards the
fixed point along those directions associated with the stable
modes.

 A nonlinear wavetrain solution
should appear as a periodic orbit, but for clarity we represent it as a fixed
point in Fig.~7; our argument is
unchanged. The entire family of these nonlinear wave solutions plots out a
semi-infinite, one-dimensional curve in the state space, which
we parametrise by $\lambda$, the wavelength. The foot of this curve is the homogeneous state.
 In Fig.~7 the thick curve represents this continuous family of
solutions, and the thinner arrowed curves represent possible trajectories of
the system.

\begin{figure}[!ht]
\begin{center}
\scalebox{.5}{\includegraphics{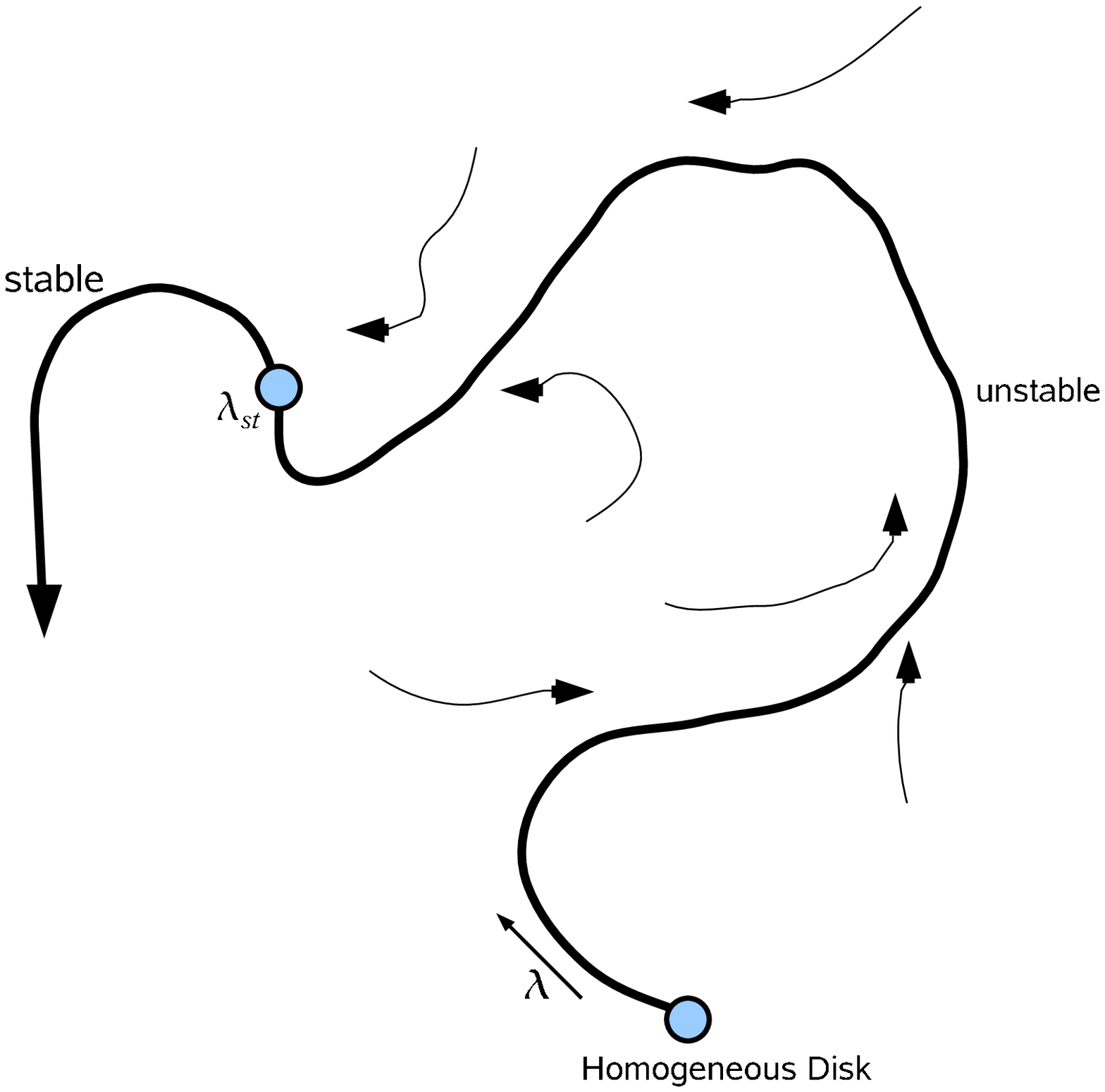}}
\caption{\footnotesize{A schematic drawing showing a two-dimensional
    projection of the overstable disk's state space, and the upward cascade to
    longer lengthscales.
  The thick curve
    represents the continuous family of nonlinear wavetrains, parametrised by $\lambda$,
  and the thin curves represent typical trajectories. Also plotted are the
  state of homogeneous Keplerian shear and the first stable nonlinear
  wavetrain, for which $\lambda=\lambda_\text{st}$. The basic idea
  is that trajectories repeatedly pass
from the overstable manifold of one fixed point to the stable manifold of a
neighbouring longer wavelength fixed point. This process ends, or at least
  undergoes a qualitative change, once the
  trajectory enters the vicinity of the
  first stable fixed point.}}
\end{center}
\end{figure}

Waves are unstable if $\lambda<\lambda_\text{st}$,
 where $\lambda_\text{st}$ is some critical wavelength, while longer waves are
 stable, and like the overstable homogeneous state, the unstable wave
 solutions are saddles, each possessing a slow overstable manifold and a fast
 stable manifold. Trajectories near the
 unstable segment of the curve of fixed points will be drawn towards the curve
 along the nearest stable manifold; at the same time trajectories are
 directed `up'
 the curve by the overstable manifold. Because the time-scale of the attraction is
 faster than that of repulsion, trajectories will be bunched, or `focused', about
 the curve of fixed points even as they travel up it. Close to the
 branch of invariant fixed points this migration may be facilitated by the 
 the linear `translation mode' with
 $s=0$ and eigenvalue $\d_\theta \bm{Z}_0$  (see next subsection).
Thus the thick curve in Fig.~7 is a one-dimensional `spine' around which
collects a nest of
state-space trajectories. Such behaviour is typical
of dissipative systems, in which the long time dynamics is
governed by a lower dimensional subset of the state space (e.g. Robinson 2001).

The upward drift will end, or at least change qualitatively, once a
 trajectory reaches the vicinity of the first stable fixed point. 
Because larger $\lambda$ fixed points are, necessarily, associated with longer
wavelengths this upward migration should coincide with the movement of power
to longer and longer scales.

It is not immediately obvious what happens once a trajectory nears the first
stable fixed point, nor how a trajectory behaves if it begins near the stable
portion of the solution branch. 
The simplest outcome is for the system
to settle on the first stable fixed point available ($\lambda_\text{st}$ for
example), but this is far from assured.
The basin of attraction of a stable fixed point may be very small, and we
expect this to be the case for most parameters.
 It is more likely that the system will `wander
around' 
the set of stable solutions, yet never fall
 upon any particular one. Physically this `wandering' will appear
 as temporal and spatial
 modulations upon a bed of nonlinear waves, with the modulations manifesting on
 lengthscales longer than that of the underlying wavetrains.
 It is not clear which set of $\lambda$ the system
 will select for the underlying waves, or whether this set
 gradually changes (perhaps growing larger and larger with time). Moreover,
 the situation may be complicated by a chaotic attractor (as can be the case in
 the complex Ginzburg-Landau equation). That said,
it is plausible that small initial conditions will yield
 a band of saturated wavelengths near
 $\lambda_\text{st}$.

These conjectures are straightforward to check with
numerical simulations, and we are currently undertaking this work. 
The only
published simulation, that of Schmit and Tscharnuter (1999), exhibited an
upward cascade which in the nonself-gravitating case had not yet halted
after $10,000$ orbits. By that stage most power was situated on a scale of
some $100H$. According to the linear stability calculations of the
previous section, using the same parameters as Schmit and Tscharnuter, 
 we find $\lambda_\text{st} \approx 60 H$; so
the simulation definitely passed the critical wavelength of the linear
analysis.
Recall, however,
that these simulations do not support
 global nonlinear 
wavetrain solutions on account of the boundary conditions and this negates
 much of the interpretation of this section.
In their simulations, trains of travelling waves can emerge locally but given sufficient time
(and 10,000 orbits is more than enough) these pulses will traverse the domain,
encounter the
boundaries, and reflect backwards and interfere with themselves. Such nonlinear interactions
are not captured in our model and perhaps induce an injection of power to longer
scales than what we would expect.

 Finally, one should
note that in our interpretation the halting of the upward cascade does not
require self-gravity, which is an effect Schmit and Tscharnuter emphasise.
 In their
simulations its inclusion halts the upward cascade. But it is probable that
the physical mechanism is different to that we have sketched.

\subsection{Modulated wavetrains: weak shocks, sources, and cellular
dynamics}

This and the next section ascertains the basic nonlinear dynamics of wavetrain modulations.
 This will give us some handle on the behaviour of the system
once it nears the set of stable fixed points described in Fig.~7. The problem
is a difficult one, and not generally amenable to analytic techniques.
 That said, a useful result can be derived if it is assumed that the
modulations in question vary slowly in time and space. When this is the
case, the phase perturbations
 are governed by the Burgers equation, which in turn suggests that the radial domain
 may fracture into regions (`cells') of
different wavenumber bounded by two kinds of interface: `weak
shocks' and `weak sources', where the phase undergoes rapid change.
The actual dynamics is perhaps even more complicated than this
ideal picture  but it remains a useful paradigm to understand the larger-scale 
irregular variations observed. 
It is also worth remarking that the theory is akin to the attractive idea proposed by Tremaine
(Araki and Tremaine 1986, Tremaine 2003) whereby disk structure corresponds to
`jams' --- the key difference is that in the Tremaine model the jams
correspond to discontinuities in shear, and in our model the jams correspond to
discontinuities in the wavenumber of background waves. 

We now very briefly describe the derivation of the Burgers equation. The proof is
lengthy and tedious so more details can be
found in Appendix B. It is essentially a multiple-scales analysis which
applies the methodology of Howard and Kopell (1977) and Doelman et al.~(2009) who examined
modulated wavetrains in reaction--diffusion systems.

\subsubsection{The Burgers equation}

Consider a linearly stable wavetrain of $\mu$ and  $\omega=\omega(\mu)$  when
 $\mu$ is small.
 We represent it
by the vector $\bm{Z}_0(\theta\,;\,\mu)$, where we have made the
dependence of $\bm{Z}_0$ on wavenumber $\mu$ explicit
in contrast to Section 4. The wavetrain is assumed stable; therefore, associated with
it are a set of decaying linear modes possessing negative growth rates. Of these
we select the modulational viscous instability mode and denote its growth rate
by $s(k)$ where $k$ is its wavenumber. From Section 5 and Eq.~\eqref{modvisc}, this mode
always decays as $k^2$.

Slowly varying modulations of $\bm{Z}_0$ are sought. First we
establish the long length and time scales characteristic of this modulation. 
For a small dimensionless parameter
$0<\delta\ll 1$, define new (slow) variables
$$ X= \delta( x - c_g\,t), \qquad T= \delta^2\,t, $$
where $c_g$ is the group velocity of the underlying wavetrain:
$$ c_g\equiv \frac{d \omega}{d \mu}.$$
Note that we have chosen a spatial frame moving with the group velocity.
Wavetrain modulations travel at $c_g$, not the phase
velocity $c_p$.

 Next a perturbation of the wavetrains' phase is
introduced, $\Theta(X,T)$. The associated variation in
wavenumber is $\d_X\Theta$. We now construct a solution of the form
\begin{equation} \label{ansatz}
\bm{Z}=
\bm{Z}_0\left(\theta+\Theta\,;\,\mu+\delta\,\Theta_X\right) + \delta^2\,\bm{Z}_1(\theta,X,T),
\end{equation}
where a subscript $X$ (or $T$) indicates partial differentiation. Our strategy
is to derive an equation for the phase modulation $\Theta$ so that the ansatz
\eqref{ansatz} satisfies the nonlinear equations to the leading orders in $\delta$.

The ansatz \eqref{ansatz} is substituted into
\eqref{cuntnl}--\eqref{cuntnl3} and these equations are expanded in powers of
the small parameter
$\delta$. At leading order $\delta^0$ the balance becomes the nonlinear
eigenproblem for the underlying wavetrain profile, Eqs
\eqref{odes1}--\eqref{odes3}. At the next order $\delta^1$ the balance is a
simple identity: the first $\mu$ derivative of the leading order equations.
At order $\delta^2$ we obtain
\begin{equation} \label{d9}
\mathcal{L}_0\,\bm{Z}_1 = -(\d_\theta\bm{Z}_0)\,\Theta_T +
\bm{A}(\theta)\,\Theta_{XX} - \bm{B}(\theta)\,\Theta_X^2, 
\end{equation}
where $\mathcal{L}_0$ is the linear operator introduced in Section 5, 
the $\bm{A}$ and $\bm{B}$ are vectors that depend only on
the underlying wavetrain and so depend only on $\mu$ and $\theta$. Their expressions are
complicated and we omit them here (see Appendix B). The order $\delta^2$
equation need only yield a solvability condition: the right side of
Eq.~\eqref{d9} must lie in the range of $\mathcal{L}_0$ (the Fredholm alternative). This can be assured
if the inner product of the right side with the adjoint solution of $\mathcal{L}_0$ is zero.
 After
some laborious manipulation, the condition is equivalent to the Burgers equation,
\begin{equation}  \label{Burgers}
 K_T + \omega''(\mu)\, K\, K_X +\tfrac{1}{2}s''(0)\,K_{XX} =0, 
\end{equation}
where $K\equiv\Theta_X$ is the wavenumber modulation, and
$$ \omega''(\mu) \equiv \frac{d^2 \omega}{d\mu^2}, \qquad \ s''(k)\equiv
\frac{d^2 s}{d k^2}.$$ The `phase diffusion coefficient' is $-s''(0)/2$ and is associated with
the decaying modulational viscous instability. Localised disturbances which
`bunch up' the crests of wavetrains will relax according to this mode. 
 The `advective coefficient' is $\omega''$ and thus
 equal to the group velocity's rate of change
  with $\mu$. It controls the steepening of
 fronts or shocks in $K$. If $c_g$ is independent of $\mu$ there is no
 wave steepening.
 How one derives $s''(0)$ from  $\bm{A}$ and 
$\omega''(k)$ from
 $\bm{B}$ is nontrivial and is left to Appendix B. There we also show how to
 compute the coefficients in terms of the material properties of the disk,
 $\alpha$, $\alpha_b$, and $\beta$.

\subsubsection{Weak shocks and sources}

For our purposes, equation \eqref{Burgers} predicts two basic
behaviours. Localised solutions $K(X,T)$ will decay to zero as $T\to\infty$
while propagating at a speed equal to the group velocity (Whitham, 1974,
Doelman et al.~2009). This behaviour simply expresses the linear stability of
the underlying wavetrain.
Nonlocalised solutions, on the other hand, can manifest as viscous Lax
shocks. For instance, suppose that we require
\begin{align*}
& K\to K_+ \quad \text{as}\quad X\to\infty, \\
& K\to K_- \quad \text{as} \quad X\to-\infty,
\end{align*} 
 where $K_+$ and $K_-$ are two real
constants. Then Eq.~\eqref{Burgers} admits a travelling front or shock
solution $ K(X,T)= K_\text{sh}(X- c_\text{sh}\,T)$ when
\begin{equation} \label{shock}
   \omega''(\mu) (K_+ - K_-)<0.
\end{equation}
From Fig.~4 or from Eqs \eqref{nldispt} and \eqref{HHt} we have $\omega''>0$
and thus $K_+<K_-$. In addition, the speed of the front is given by
$$ c_\text{sh} = \tfrac{1}{2} \omega''(\mu)\,(K_+ + K_-),$$
which is the Rankine-Hugoniot condition in the moving frame.
The characteristics associated with the solution point towards the front
interface, as do the motions of the individual wavecrests on its each side.
 Solutions for which the characteristics and wavecrests point away do not
 satisfy \eqref{shock} and
correspond to rarefaction waves or `sources'. Fig.~8 presents a cartoon of a
nonlinear wavetrain exhibiting
both a shock and a source at which points the wavenumber (and amplitude)
rapidly change.

\begin{figure}[!ht]
\begin{center}
\scalebox{.7}{\includegraphics{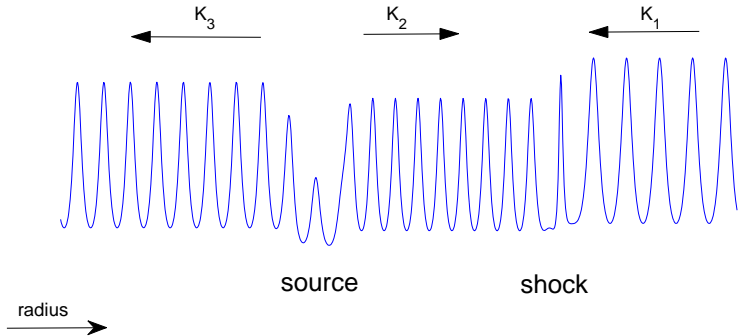}}
\caption{\footnotesize{A simple cartoon representing a wavetrain with a shock and a
 source solution. The arrows represent the direction that the
 individual wavecrests propagate (the $c_p$), while $K$ represents the
modulational wavenumber. Here $K_1<K_3<K_2$ are constants, and thus the 
defects `glue' together patches of different wavenumber (and hence amplitude).
On sufficiently large scales a wavetrain may decompose into a number of
 subtrains each possessing varying wavenumber and each
 enclosed by a shock and a source. These interfaces will possess
 their own slow dynamics --- they may wander around the domain, 
annihilate one another, generate new defects, etc. }}
\end{center}
\end{figure}

\subsubsection{Cellular structure}

In general reaction-diffusion systems, widely-spaced sources and shocks
can partition the domain into distinct regions inhabited by wavetrains of different
$\mu$ (Aranson and Kramer 2002, Doelman et al.~2009). Figure 8 roughly sketches a
portion of a domain so decomposed. Similar phenomena should emerge in
overstable disks, and it is possible that observed
radial structure on intermediate scales in Saturn's rings corresponds to the
cellular patterns constructed from such wave defects.

Shock and souce interfaces will possess their own slow
dynamics which compels them to drift relative to each other and undergo various
kinds of interactions. The former motions may be represented by a
 simple one-dimensional `cellular' dynamical
 system that evolves the interfaces according to a set of ordinary
 differential equations (Elphick et al.~1990,
 Ei 2002, Doelman et al.~2009). Each interface is treated as a particle that weakly
 interacts with its neighbours via a `potential' and
 thus the model bears a similarity to a one-dimensional $N$-body problem.

 Unfortunately, it is difficult to produce analytic estimates for the
 characteristic lengthscales of the cellular
 structures so formed. Thus comparison with Cassini data must wait.
 In any case, the
 model assumes only weak interactions between the interfaces, but full
 numerical simulations of the complex Ginzburg-Landau equation typically exhibit
 more complicated behaviour (Popp et
 al.~1993, Chat\'{e} 1994). We briefly survey these phenomena in the next subsection.

\subsubsection{The complex Ginzburg-Landau equation}

The complex Ginzburg-Landau
equation in an extended domain boasts a substantial literature, equal to the
diverse behaviour it exhibits. This body of work may guide the
interpretation of future nonlinear simulations of the overstability.
The complex Ginzburg-Landau system crystallises the essential dynamics of general
nonlinear wave phenomenon in a broad range of physical situations (Aranson and
Kramer 2002) and which we expect nonlinear waves in planetary rings to share.\footnote[1]{
 In fact, the complex Ginzburg-Landau equation
has been derived from a weakly nonlinear analysis of the overstability in a
simple hydrodynamic model (J.~Schmidt, private communication).} 

 The one-dimensional
version of the equation reads
$$ \d_t A =A + (1+i c_1)\d_x^2 A -(1+i c_2)|A|^2\,A,$$
where $A$ is a complex function and $c_1$ denotes the linear dispersion and $c_2$
nonlinear dispersion. The function $A$ typically represents the complex
amplitude of some wave phenomenon, and is not the same as the phase modulation
$\Theta$ which we studied earlier --- the latter, though, can be derived from $A$.
The equation gives rise to steady, nonlinear wavetrains,
various wave instabilities, and `defects' (such as source and shock
structures) that can `glue' together different wavetrains,
 in addition to different types of spatio-temporal chaos
(Aranson and Kramer 2002). 

Of the various regimes exhibited, behaviour
associated with the
so-called `intermittency regime' is of most interest.
 Here the state space admits not only stable
nonlinear wavetrain solutions but also a `defect-turbulent attractor'
(Chat\'{e} 1994). The role of the latter is to govern the irregular dynamics 
of the shocks and sources that separate the different regions of stable
travelling waves.
 A number of studies have explored the rich and chaotic dynamics of
 the interfaces (Popp et al.~1993, Chat\'{e} 1994, Ipsen and Hecke 2001): 
for instance,
source and shock structures may wander around the domain,
 annihilate each other, and give birth to new shock and source pairs. The 
long-time outcome is also variable, and depends closely on the parameters. For
certain choices of $c_1$ and $c_2$, the final state may be `glassy' (a static
arrangement of defects), the defects may evaporate altogether, or the birth
and annihilation process continues indefinitely. Another interesting regime
is that of `phase chaos' in which the wavenumber of the nonlinear
wavetrains undergo turbulent fluctuations. All or some of this behaviour we
expect to carry over to the dynamics of the overstable disk. At the very
least, the complex Ginzburg-Landau equation provides a template
with which to interpret simulations of overstable disks 
and the ISS observations of irregular structure in the B-ring.

\section{Conclusion}

This paper has investigated the dynamics of an overstable fluid disk
by exploiting the properties of its nonlinear invariant solutions.
 These structures manifest as
axisymmetric uniform travelling wavetrains and form a continuous family of
 solutions parametrised by their wavelength $\lambda$. In Section 4 and
 Appendix A we computed the members of this family and outlined their physical
 characteristics. In particular, we find that inertial forces dominate their
 leading order force balance, and so the fluid undergoes approximately epicyclic
 motion. The wave amplitudes, on the other hand, are determined from an energy balance that counterpoises
  linear viscous
 overstability and viscous dissipation.
 In Section 5
 the wavetrains' stability to small perturbations is inspected. It turns out
 that only
 solutions with wavelengths above a critical value $\lambda_\text{st}$ are
 linearly stable. If the viscous parameters take values associated with
 Saturn's B-ring,  $\lambda_\text{st}\approx 50-100\,H$. Finally, in Section 6 
we speculate on the role of the nonlinear wavetrains in the general dynamics,
arguing that the system will linger `near' the branch of linearly stable invariant
solutions. Physically, this phase-space behaviour will translate to
a bed of nonlinear waves suffering large-scale modulations or defects,
the latter partitioning the domain into a cellular grid of wavetrains of
different $\lambda$.

 Our basic hypothesis is that quasi-periodic
microstructure on scales of order $100$m revealed by the UVIS and RSS correspond to the
underlying bed of nonlinear waves. Meanwhile,
 irregular structure on larger scales, of 500m to 10km may be
 a manifestation of the wavetrains' spatio-temporal
modulations, arising as the system flirts with a number of stable wavetrain
solutions but is unable to settle on any.
 The individual waves are beneath the ISS cameras' resolution, but
 larger-scale variations in their amplitude and wavelength
 should give rise to different optical properties which
the cameras can register. In order to fully establish the latter point, however, 
the optical properties of the nonlinear waves need to be
understood. The dynamical and photometric modelling techniques that have been developed 
to understand the observational
properties of gravitational wakes could be applied to this end (French et
al.~2007, Colwell et al.~2007).

This study has adopted some rather strong modelling assumptions and
neglected some important physics. The issue requiring the most urgent
attention is self-gravity, as it will certainly play a role in the formation and
stability of the nonlinear waves studied. Its effects we examine
in a following paper. The 1D isothermal fluid model we have 
employed is simple, but we doubt that the basic physics it has revealed
is an artefact of its simplicity. That said, on at least two counts it can be
improved: first, its treatment of the viscous stress, and second, its
neglect of vertical motion. The Navier--Stokes stress prescription is not
ideal in a scenario of densely-packed and infrequently colliding particles,
where the collisional stress dominates. A
detailed kinetic treatment is perhaps the best way to capture the nonlinear 
dependence on strain associated with a collisional stress (Latter and
Ogilvie 2008). Vertical motions, such as `breathing' or `splashing',
 certainly accompany both the linear modes of
the viscous overstability and its nonlinear development (Salo et al.~2001, Latter and Ogilvie
2006b). The importance of this effect is difficult to judge; and may only be
resolved by 2D eigensolutions and simulations, both formidable tasks.

Lastly, we briefly sketch out future work. Once the role of self-gravity on
the nonlinear waves has been established, large-scale hydrodynamical 
 simulations of an overstable ring must be undertaken to test our
various conclusions. These should also address the influence of the boundary
conditions, which will help us connect the work here to that
 of Schmit and
Tscharnuter (1999). On the other hand,
 the dense--gas kinetic equations may be simulated, using
perhaps the continuum model derived by Latter and Ogilvie (2008). These then
could provide a more direct comparison with $N$-body simulations and possibly
to the observations of finescale structure themselves.

\section*{Acknowledgements}

The authors thank the two reviewers, J\"{u}rgen Schmidt and Frank Spahn, for
their careful and thorough reading of the manuscript. It has benefited greatly
from their time and effort. This research was partly funded via an STFC grant.

\section{Appendix A: Asymptotic description of ring dynamics in the limit of small
  collective effects}
The analysis presented in this appendix resembles earlier work on nonlinear waves in disks.
  Several papers by Shu and collaborators formulate equations for
nonlinear spiral density waves in planetary rings and accretion discs (the
most pertinent being Shu et al.~1985a, 1985b).  They include
self-gravity, pressure and viscous (or kinetic) stresses.  However, they do not
consider time-evolution or viscous overstability because they are interested in
forced rather than free waves. The streamline formalism of Borderies et al.\
(1983, 1986) is closely related.  They derive their
equations using methods of celestial mechanics but the outcome is similar.  Because
they consider discrete streamlines rather than a continuous medium, they obtain
coupled ODEs rather than PDEs which means that density waves are not so clearly
manifested in their approach. Our analysis is intentionally restricted to
axisymmetry and local geometry, which slightly clarifies matters and allows
viscous overstability to take centre stage. The physics of self-gravity, nonaxisymmetry, and
the gravitational influence of external moons may be included later.

\subsection{Basic equations}

We consider axisymmetric motions in a shearing sheet with no vertical
extension.  The sheet is Keplerian and units are chosen such that
$\Omega=1$.  The basic equations are
\begin{equation}
  \rmD u-2v=\epsilon T_x,
\end{equation}
\begin{equation}
  \rmD v+\f{1}{2}u=\epsilon T_y,
\end{equation}
\begin{equation}
  \rmD\sigma+\sigma\p_xu=0,
\end{equation}
where $\rmD=\p_t+u\p_x$ is the Lagrangian derivative, $T_x$ and $T_y$ are
the radial and azimuthal accelerations due to the weak collective effects of
pressure, viscosity, and self-gravity, and
 $\epsilon$ is an ordering parameter. Note that the notation here is different
 to the main body of the paper: $u$ denotes the radial velocity and $v$ the
 azimuthal velocity.

We anticipate that the dominant motion of fluid elements is an epicyclic oscillation of
the form
\begin{equation}
  x=\xi+a(\xi,t)\cos[t-b(\xi,t)],
\end{equation}
where $a$ and $b$ are the epicyclic amplitude and phase, and $\xi$ is
an \emph{approximate} Lagrangian variable.  It is not an exact Lagrangian
variable because the exact motion of a fluid element in a viscous disk departs
marginally from epicycles.  In fact, we use this equation to
\emph{define} $\xi(x,t)$, and then determine how $a$ and $b$ must evolve so that
this equation continues to describe the dominant motion of the
fluid.  This constraint is meaningful only if $a$ and $b$ evolve slowly in
time, a constraint enforced by the asymptotic analysis below.
  We write $a_\xi$ and $a_t$ to represent $\p a/\p\xi$ and $\p a/\p t$,
 respectively.  We also work with the complex epicyclic amplitude
\begin{equation}
  A=ae^{ib}.
\end{equation}
The relation between $x$ and $\xi$ is one-to-one provided that $|A_\xi|<1$.
Interpenetration of neighbouring fluid elements (or streamlines) occurs if $|A_\xi|>1$.

We now change variables from $(x,t)$ to $(\xi,t)$, finding
\begin{equation}
  \rmD u-2v=\epsilon T_x,
\label{du}
\end{equation}
\begin{equation}
  \rmD v+\f{1}{2}u=\epsilon T_y,
\label{dv}
\end{equation}
\begin{equation}
  \rmD\sigma+\sigma J^{-1}\p_\xi u=0,
\label{dsigma}
\end{equation}
where now
\begin{equation}
  \rmD=\p_t+\tilde uJ^{-1}\p_\xi,
\end{equation}
\begin{equation}
  \tilde u=u-\dot x,
\end{equation}
\begin{equation}
  J=\left(\f{\p x}{\p\xi}\right)_t=1+a_\xi\cos(t-b)+ab_\xi\sin(t-b),
\end{equation}
\begin{equation}
  \dot x=\left(\f{\p x}{\p t}\right)_\xi=a_t\cos(t-b)-a(1-b_t)\sin(t-b).
\end{equation}
(Note that $\p_t$ now denotes a derivative at constant $\xi$.)  $J$ is the
Jacobian (stretching factor) of the transformation between $x$ and $\xi$,
while $\dot x$ is the radial velocity of a particle having constant $\xi$.  If
$\xi$ were an exact Lagrangian variable, then $u$ would equal $\dot x$,  
therefore $\tilde u$ is the radial velocity relative to the moving coordinate system.  Note the identity
\begin{equation}
  \p_tJ=\p_\xi\dot x.
\label{id}
\end{equation}

\subsection{Asymptotic solution}

We propose that $a$ and $b$ evolve, due to weak collective effects, on a timescale that is long (by a
factor of order $\epsilon^{-1}$) compared to the orbital timescale.
Thus $a$ and $b$ are considered as functions of $(\xi,\tau)$, where
$\tau=\epsilon t$ is a slow time variable, and $a_t$ becomes $\epsilon a_\tau$, etc.

The solution is expanded in the form
\begin{align*}
  \tilde u &=\epsilon\tilde u_1+\cdots, \\
  v &=v_0+\epsilon v_1+\cdots, \\
  \sigma &=\sigma_0+\epsilon\sigma_1+\cdots, \\
\end{align*}
where $\tilde u_1$, etc., are functions of $(\xi,t,\tau)$. (Note that these
functions are different to those appearing in Section 5.1 though the notation
is the same.)  The multiple-scale analysis is necessary in order to separate the fast epicyclic oscillation from the slow evolution of the epicyclic amplitude and mass distribution.  All the variables should be periodic functions of $t$ with period $2\pi$; any secular evolution is described by the dependence on $\tau$.

Note that $a$ and $b$ are not expanded in powers of $\epsilon$, because they
represent only the leading-order motion.  Thus $J(\xi,t,\tau)$ is a quantity of
order unity and does not require expansion.  On the other hand $\dot x$ contains terms of orders $\epsilon^0$ and $\epsilon^1$.  The relation between $u$ and $\tilde u$ is
\begin{equation*}
  u-\tilde u=\dot x=-a\sin(t-b)+\epsilon[a_\tau\cos(t-b)+ab_\tau\sin(t-b)]=\dot x_0+\epsilon\dot x_1.
\end{equation*}
The identity (\ref{id}) becomes
\begin{equation}
  \p_tJ=\p_\xi\dot x_0,\qquad
  \p_\tau J=\p_\xi\dot x_1.
\label{id2}
\end{equation}

At leading order we have the epicyclic motion
\begin{equation}
  \dot x_0=-a\sin(t-b),\qquad v_0=-\f{1}{2}a\cos(t-b).
\end{equation}
This satisfies the equations of motion (\ref{du}) and (\ref{dv}) at $O(1)$.
The equation of mass conservation (\ref{dsigma}) at $O(1)$ gives
\begin{equation}
  \p_t\sigma_0+\sigma_0J^{-1}\p_\xi\dot x_0=0.
\end{equation}
Multiplying by $J$ and using the identity (\ref{id2}), we obtain
\begin{equation}
  \p_t(\sigma_0J)=0,
\end{equation}
which expresses conservation of mass on the epicyclic timescale.  We may write
 $\sigma_0J=\Sigma(\xi,\tau)$. 

 Our task now is to deduce evolutionary equations for the epicyclic amplitude
 $A(\xi,\tau)$ 
and the epicycle-averaged surface density $\Sigma(\xi,\tau)$.
  (Note that, since $dx=J\,d\xi$ at fixed $t$, $\Sigma$ is really a density with respect to the $\xi$ coordinate.)

\subsubsection{Evolution equations}

The equation of mass conservation at $O(\epsilon)$ is
\begin{equation}
  \p_t\sigma_1+(\p_\tau+\tilde u_1J^{-1}\p_\xi)\sigma_0+\sigma_0J^{-1}\p_\xi(\tilde u_1+\dot x_1)+\sigma_1J^{-1}\p_\xi\dot x_0=0.
\end{equation}
Multiplying by $J$ and using the identity (\ref{id2}), we find
\begin{equation*}
  \p_t(\sigma_1J)+\p_\tau(\sigma_0J)+\p_\xi(\sigma_0\tilde u_1)=0.
\end{equation*}
We now apply the epicyclic averaging operation
\begin{equation*}
  \langle\cdot\rangle=\f{1}{2\pi}\int_0^{2\pi}\,\cdot\,dt,
\end{equation*}
to obtain
\begin{equation}
  \p_\tau\Sigma+\p_\xi(\Sigma V)=0.
\end{equation}
where $V(\xi,\tau)$ is a mean radial velocity defined by the averaged mass flux density
\begin{equation}
  \Sigma V=\langle\sigma_0\tilde u_1\rangle=\Sigma\langle\tilde u_1J^{-1}\rangle.
\end{equation}
What we have done here is to obtain an evolutionary equation for $\Sigma$ by applying a solvability condition that eliminates the unknown $\sigma_1$.

The equations of motion at $O(\epsilon)$ are
\begin{align}
  &\p_t(\tilde u_1+\dot x_1)+(\p_\tau+\tilde u_1J^{-1}\p_\xi)\dot x_0-2v_1=T_x,\\
  &\p_tv_1+(\p_\tau+\tilde u_1J^{-1}\p_\xi)v_0+\f{1}{2}(\tilde u_1+\dot
  x_1)=T_y. 
\end{align}
Our aim is to eliminate the unknowns $\tilde u_1$ and $v_1$ by obtaining a solvability condition that will provide an evolutionary equation for the epicyclic amplitude $A$.  We will show how this can be obtained in a systematic manner.  Let us rewrite the equations, noting that $\p_\xi v_0=(1-J)/2$ and $\p_\tau v_0=-\dot x_1/2$, in the form
\begin{align*}
  \p_t\tilde u_1+\f{\tilde u_1}{J}\p_tJ-2v_1 &=\tilde T_x \,\equiv\,T_x-\p_t\dot x_1-\p_\tau\dot x_0,\\
  \p_tv_1+\f{\tilde u_1}{2J} &=T_y.
\end{align*}
The latter (angular momentum) equation, when averaged, gives us immediately
\begin{equation}
  V=2\langle T_y\rangle.
\end{equation}
Now eliminate $\tilde u_1$ between the two equations to obtain
\begin{equation*}
  -2Jv_1-2\p_t(J^2\p_tv_1)=J\tilde T_x-2\p_t(J^2T_y)
\end{equation*}
On the left-hand side we have a linear operator in self-adjoint form acting on $v_1$.  It can be verified that $J^{-1}e^{it}$ is a null eigenfunction.  The corresponding solvability condition is
\begin{equation}
  \langle J^{-1}e^{it}[J\tilde T_x-2\p_t(J^2 T_y)]\rangle=0.
\end{equation}
After an integration by parts and use of the identity
\begin{equation*}
  i\p_tJ=1-J+A_\xi e^{-it}
\end{equation*}
we have
\begin{equation*}
  \langle e^{it}\tilde T_x+2i(e^{it}+A_\xi)T_y\rangle=0,
\end{equation*}
and so
\begin{equation}
  \p_\tau A+V\p_\xi A=\langle e^{it}(iT_x-2T_y)\rangle.
\end{equation}
In summary, we have derived the evolutionary equations
\begin{equation} \label{asc1}
  \p_\tau\Sigma+\p_\xi(\Sigma V)=0,
\end{equation}
\begin{equation}\label{asc2}
  \p_\tau A+V\p_\xi A=\langle e^{it}(iT_x-2T_y)\rangle,
\end{equation}
with
\begin{equation} \label{asc3}
  V=2\langle T_y\rangle.
\end{equation}
The evolution of the mass distribution and the nonlinear waves are coupled
 together. 
 We can now remove the ordering parameter and replace $\p_\tau$ by $\p_t$ here
 if desired. 
 However, it is important to retain the conceptual distinction between the fast and slow timescales.

\subsection{Evaluation of the force averages}

\subsubsection{Stress tensors in general}

If the forces derive from a (two-dimensional/vertically integrated)
stress tensor, then
\begin{equation*}
  T_x=-\f{1}{\sigma}\p_x f,\qquad
  T_y=-\f{1}{\sigma}\p_x h,
\end{equation*}
where $f$ and $h$ are as given in Sections 3 and 4 in the case of a fluid with
an isotropic pressure and a Navier-Stokes viscous stress. Self-gravity is omitted.
To calculate the stresses to leading order, 
we replace $\sigma$ with $\sigma_0$ here and use the leading-order velocity field where required.
Also, noting that $\sigma_0^{-1}\p_x=\Sigma^{-1}\p_\xi$ and that $\Sigma$ is independent of the fast time $t$, we have
\begin{equation}
  \langle e^{it}(iT_x-2T_y)\rangle=-\f{1}{\Sigma}\p_\xi\langle e^{it}(if-2h)\rangle,
\end{equation}
\begin{equation}
  \langle T_y \rangle=-\f{1}{\Sigma}\p_\xi\langle h\rangle.
\end{equation}

\subsubsection{Pressure}

We first compute the pressure contribution to the evolutionary equations by
setting the viscous stress to zero.
In the case of an isothermal gas with (two-dimensional/vertically integrated)
pressure $p=v_s^2 \sigma$, we have
\begin{equation}
  f= v_s^2\Sigma J^{-1},\qquad h=0.
\end{equation}
We must then evaluate $\langle
e^{it}J^{-1}\rangle$.  Let
\begin{equation}
  A_\xi=(a_\xi+iab_\xi)e^{ib}=qe^{i\phi}.
\end{equation}
Here $0\le q<1$ is the usual nonlinearity parameter for density waves in
planetary rings (see Borderies, Goldreich, and Tremaine 1983, 1985, 1986).
Then
\begin{equation*}
  J=1+\mathrm{Re}[A_\xi e^{-it}]=1+\mathrm{Re}[qe^{-i(t-\phi)}]=1+q\cos(t-\phi)=1+q\cos\theta
\end{equation*}
and
\begin{equation*}
  \langle e^{it}J^{-1}\rangle=e^{i\phi}\langle e^{i\theta}(1+q\cos\theta)^{-1}\rangle=e^{i\phi}\f{1}{2\pi}\int_0^{2\pi}\cos\theta(1+q\cos\theta)^{-1}\,d\theta,
\end{equation*}
where $\theta=t-\phi$. This is straightforward to integrate and the
evolutionary equations with pressure alone become
\begin{equation}
  \p_t\Sigma=0,
\end{equation}
\begin{equation}
  \p_tA=\frac{i v_s^2}{\Sigma} \p_\xi\left[ F_1(q)\,\Sigma\, A_\xi\right],
\end{equation}
with
\begin{equation} \label{FF1}
F_1(q)= q^{-2}\,\left[(1-q^2)^{-1/2}-1\right].
\end{equation}
This is a type of nonlinear Schr\"odinger equation.  Note that $q^2=|A_\xi|^2$.

\subsubsection{Viscosity}
Next we compute the contribution from viscosity alone, in the absence of
pressure. Then
\begin{equation}
  f=-(\nu_b+{\textstyle\f{4}{3}}\nu)\sigma\p_xu,\qquad
  h= -\nu\sigma(-{\textstyle\f{3}{2}}+\p_xv),
\end{equation}
and we adopt the viscosity laws
\begin{equation}
  \nu=\frac{v_s^2}{\Omega}\,\alpha\left(\f{\sigma}{\sigma_*}\right)^\beta,\qquad
  \nu_b=\frac{v_s^2}{\Omega}\,\alpha_{b}\left(\f{\sigma}{\sigma_*}\right)^\beta,
\end{equation}
where $\sigma_*$ is a constant reference density.
To leading order and in terms of the variables $Q=qe^{i\phi}$ and $\theta=t-\phi$
introduced previously,
\begin{equation}
  f= v_s^2\,(\alpha_{b}+{\textstyle\f{4}{3}}\alpha)\sigma_*\left(\f{\Sigma}{\sigma_*}\right)^{\beta+1}J^{-\beta-2}q\sin\theta,
\end{equation}
\begin{equation}
  h= v_s^2\, \alpha\sigma_*\left(\f{\Sigma}{\sigma_*}\right)^{\beta+1}J^{-\beta-2}({\textstyle\f{3}{2}}+2q\cos\theta).
\end{equation}
The weighted averages of these relations, which appear in Eqs
\eqref{asc1}--\eqref{asc3}, are complicated integral functions which may be
rewritten in terms of associated Legendre functions. In summary,
\begin{align}
&\langle e^{it}(if-2h)\rangle = -\sigma_* v_s^2\,
  \left(\frac{\Sigma}{\sigma_*}\right)^{1+\beta} F_2(q)\,A_\xi, \\
&  \langle h\rangle = -\sigma_* v_s^2\,
  \left(\frac{\Sigma}{\sigma_*}\right)^{1+\beta} F_3(q),
\end{align}
where we have introduced two new $F$ functions defined through
\begin{align}
F_2(q) &= \frac{(\widetilde{q})^{2+\beta} }{\beta(1+\beta)\,\, q } \left\{ (\alpha_b+\tfrac{16}{3}\alpha)\frac{1}{\widetilde{q}}\,\text{P}^{(1)}_\beta(\widetilde{q})
- 3 \alpha\beta\,\text{P}^{(1)}_{1+\beta}(\widetilde{q}) +
4\alpha\,q\,\text{P}^{(2)}_{1+\beta}(\widetilde{q})  \right\}   \label{FF2} \\
F_3(q) &= - \frac{\alpha\,\widetilde{q}^{2+\beta}}{1+\beta}\,\left\{
  3(1+\beta)\,\text{P}^{(0)}_{1+\beta}(\widetilde{q}) -
  4q\,\text{P}^{(1)}_{1+\beta}(\widetilde{q})  \right\}, \label{FF3}
\end{align}
where  $\widetilde{q}=(1-q^2)^{-1/2}$, a notational convenience, and
$\text{P}^{(m)}_{\eta}$ is the associated Legendre function of the first kind and type
3. It is defined through
$$ \text{P}^{(m)}_\eta(z)= \frac{\Gamma(m-\eta)}{\pi\Gamma(\eta)}\int_0^\pi
\cos(m\theta)\left[z+\sqrt{z^2-1}\cos\theta\right]^{-\eta-1}\,d\theta $$
(see http://functions.wolfram.com/HypergeometricFunctions/LegendreP3General).

\subsubsection{Combined evolutionary equations}

Dimensions are adopted so that $v_s=1$ and $\sigma_*=1$. Finally,
 we arrive at a problem of the form
\begin{equation} \label{fina1}
  \p_t\Sigma+\p_\xi(\Sigma V)=0,
\end{equation}
\begin{equation} \label{fina2}
  \p_t A+V\p_\xi A= \f{1}{\Sigma}\,\p_\xi\left\{\left[i  F_1(q)+
  \Sigma^{\beta}F_2(q)\right]\, \Sigma\, A_\xi\right\},
\end{equation}
where
\begin{equation} \label{fina3}
  V=  \f{1}{\Sigma}\p_\xi\left[ \Sigma^{1+\beta} F_3(q)\right],
\end{equation}
and where the $F_i$'s are the real functions presented in Eqs \eqref{FF1} and
\eqref{FF2}--\eqref{FF3}. 

The behaviour of the $F_i$ controls to a large
extent the evolution of the system. When solutions are of small amplitude and
$q$ is near 0, the three $F$ functions are approximately constant because $\text{P}^{(m)}_\eta(\widetilde{q})
\propto q^{m}$. In
particular
$F_3(0)=-3\alpha$ and 
\begin{equation}
  F_2(0)= \tfrac{1}{6}\left[ 3\alpha_b-2\alpha-9\alpha\beta \right].
\end{equation}
So, we have linear viscous overstability (antidiffusion) when $F_2(0)<0$ (cf.\ Eq.\eqref{volin}),
and viscous diffusion when $F_2(0)>0$. On the other hand, if the disk is
overstable then $F_2$ must change sign as $q$ increases because when $q\to 1$
$F_2$ is positive. This suggests that
the instability saturates once the solution is sufficiently nonlinear. The
function $F_3$ also changes sign at a critical value of $q$, a situation which corresponds
to angular momentum flux reversal. It appears, for all realistic
parameters, however, that viscous overstability saturates at a $q$ below the
critical value of flux reversal.
In the opposite limit of $q\to 1$, $F_2$ diverges as
$(1-q^2)^{-\beta-3/2}$ because viscous diffusion is trying to combat the imminent crossing of
neighbouring streamlines.

\subsection{Nonlinear travelling waves}

The governing set of equations \eqref{fina1}--\eqref{fina3} appears rather
complicated, nevertheless it admits exact plane wave solutions
 of the form
\begin{equation*}
  A=A_0e^{i\mu\xi-i\omega t},
\end{equation*}
where $\mu$ and $\omega$ are real so that $q=q_0=|\mu A_0|=\mathrm{constant}$.
The solution also requires $\Sigma=\mathrm{constant}$ and $V=0$.
These represent uniform travelling waves in a homogeneous disk and provide an
asymptotic description of the nonlinear waves studied in the main body of the paper.

Steady wavetrains require
\begin{equation} \label{asomega}
  -i\omega=-\mu^2[i F_1(q_0)+ F_2(q_0)],
\end{equation}
which is possible only if $F_2(q_0)=0$.
  Therefore the amplitude of the nonlinear waves
(in the sense of $q$) corresponds to the condition of marginal
overstability.  We have $q_0=q_c$, where $q_c$ is the critical $q$ at which
$F_2(q)$ reverses sign. 
 (If $F_2(0)>0$, there is no linear
overstability and no travelling wave.) However, the amplitude in the sense of
$A_0$ is not limited. The nonlinear dispersion relation fixing $\omega$ is
consequently controlled by pressure: $\omega= \mu^2\, F_1(q_0)$.

\subsubsection{Linear stability}

Consider small perturbations upon the uniform travelling wave solution just described.
The small perturbations are denoted by $\delta \Sigma$, $\delta A$, etc.
Their linearised equations are
\begin{align}
 & \p_t\delta\Sigma+\Sigma\p_\xi\delta V=0, \\
 & \p_t\delta A+\delta V\p_\xi
  A=\left(\f{dF_2}{dq^2}+i\f{dF_1}{dq^2}\right)\p_\xi(\delta
  q^2A_\xi)\notag \\ & \hskip4cm  -\f{\delta\Sigma}{\Sigma}iF_1 A_{\xi\xi}+i
 F_1\p_\xi\left(\f{\delta\Sigma}{\Sigma}A_\xi\right)+iF_1\delta A_{\xi\xi}, \\
&  \delta V=\f{dF_3}{dq^2}\p_\xi\delta q^2+(\beta+1)F_3\p_\xi\left(\f{\delta\Sigma}{\Sigma}\right),
\end{align}
with
\begin{equation*}
  \delta q^2=\delta A_\xi A_\xi^*+A_\xi\delta A_\xi^*.
\end{equation*}
Solutions are of the form
\begin{align*}
  \delta\Sigma &=\Sigma(c_1E+c_1^*E^*), \\
  \delta V &=c_2E+c_2^*E^*, \\
  \delta A &=A(c_3E+c_4^*E^*),
\end{align*}
where
\begin{equation*}
  E=\exp(st+ik\xi),
\end{equation*}
$s$ is a complex growth rate, $k$ is a real wavenumber and the $c_i$ are
complex coefficients.  This allows us to write
\begin{align*}
  \delta
 &
 q^2=q^2\left\{\left[c_3\left(1+\f{k}{\mu}\right)+c_4\left(1-\f{k}{\mu}\right)\right]E
 \right.\\
& \hskip4cm \left.+\left[c_3^*\left(1+\f{k}{\mu}\right)+c_4^*\left(1-\f{k}{\mu}\right)\right]E^*\right\}.
\end{align*}
And now we have a fourth order algebraic system for the coefficients $c_i$,
\begin{align}
  & sc_1= - ikc_2, \\
  & c_2  =ik\hat F_3
  \left[c_3\left(1+\f{k}{\mu}\right)+c_4\left(1-\f{k}{\mu}\right)\right]+ik(\beta+1)F_3\,
  c_1, \\
  & (s-i\omega)c_3+i\mu c_2  =-(\hat F_2+i\hat F_1)(\mu+k)[c_3(\mu+k)+c_4(\mu-k)]
 \notag \\
& \hskip4.8cm  -i F_1\mu k c_1-i(\mu+k)^2 F_1 c_3, \\
&  (s+i\omega)c_4-i\mu c_2=-(\hat F_2-i\hat F_1)(\mu-k)[c_3(\mu+k)+c_4(\mu-k)]
  \notag \\
& \hskip4.8cm  -i F_1\mu k c_1+i(\mu-k)^2 F_1 c_4,
\end{align}
where the $F_i$ are evaluated at $q_0$ and $\hat F_i=q^2\,dF_i/dq^2$ evaluated
  at $q_0$.
  A complicated cubic dispersion relation for $s(k)$ follows.
  It has the property that $s/\mu^2$ is a (triple-valued) function of $k/\mu$
  only,
 with parameters $\alpha$, $\alpha_b$, and $\beta$, which set the viscous
  properties of the fluid and consequently $q_0$.

 In the long wavelength limit of the perturbations $|k|\ll|\mu|$, the three roots behave according to
\begin{align}
  s&=-2\hat F_2 \mu^2\,\,+\,\,O(k), \\
  s&=(\beta+1)F_3 k^2\,\,+\,\,O(k^3), \label{modvisc} \\
  s&=-2iF_1 \mu k-\f{F_1\hat F_1}{\hat F_2}k^2\,\,+\,\,O(k^3). \label{modover}
\end{align}
The first mode decays relatively rapidly, as $\alpha \mu^2$,
 on account of the fact that $F_2$ is always an
increasing function of $q^2$, and hence $\hat F_2 >0$.  The second mode is a
 generalisation of the viscous instability (cf.\ Section 3).
 For all realistic viscous parameters we have $F_3(q^2)< F_2(q^2)$, and with $F_2(q_0)=0$
 by definition, stability of this mode is assured when $\beta>-1$, the
 same criterion as for normal viscous instability. This mode controls the diffusion
 of small perturbations of wavenumber and amplitude. Lastly, the third mode
  corresponds to the generalised viscous overstability. It is stable as
 long as $F_1$ and $\hat F_1$ are both positive, which can be easily checked
 from Eq.\eqref{FF1}. In summary, all nonlinear wavetrains are stable to
 small-$k$ 
perturbations (in the asymptotic limit of long wavelengths that we are
 examining).

Numerically, these conclusions can be generalised to all $k$. We find that 
 overall stability of the modulational viscous overstability, and hence the
 wavetrain,
 can be determined by examining the
small-$k$ limit. In Section 5 in
Fig.~6 we plot the growth rate of the modulational viscous overstable mode
versus $k$ for parameters associated with $\tau=1.5$ in Table 1.

\section{Appendix B: Derivation of the Burgers equation}

In this section we provide more details of the derivation of
Eq.~\eqref{Burgers}. The proof closely follows that of Doelman et al.~(2009) 
who attack a simpler reaction---diffusion equation with a methodology developed
by Howard and Kopell (1977) and Whitham (1974). 

\subsection{Preliminaries}

The nonlinear equation governing the disk,
cf.\ Eqs~\eqref{cuntnl}-\eqref{cuntnl3}, we represent in a compact
vector form:
\begin{equation} \label{rxndif}
\d_t Z_i = D_{ij}(\sigma)\,\d_x^2 Z_j + G_i\left(\bm{Z},\d_x\bm{Z}\right),
\end{equation}
where $\bm{Z}=(\sigma,u_x',u_y')$. The diffusion matrix $D_{ij}$ is a
  function of $\sigma$ only, and possesses a first row of zeros (because the
  continuity equation has no second derivative). The nonlinear
   vector function $G_i$ depends on both $\bm{Z}$ and its
  derivative. It comprises advection, rotation, shear, pressure,
  and the density-dependence of the viscosity.

By defining $\theta$ as earlier, the equation for the nonlinear wavetrain
solutions $Z^0_i(\theta,\mu)$ is
\begin{equation}\label{rdwt}
\omega\d_\theta Z_i^0 + \mu^2\,D_{ij}\d_\theta^2 Z^0_j 
+ G_i\left(\bm{Z}_0,\mu\d_\theta\bm{Z}_0\right) =0,
\end{equation}
while the linearised problem for a small perturbation $\widetilde{Z}_i$ is
\begin{align}\label{rdlin}
\d_t \widetilde{Z}_i &= \omega \d_\theta \widetilde{Z}_i + \mu^2 D_{ij}\d_\theta^2 \widetilde{Z}_j + \mu^2
D_{ij}'\d_\theta^2 Z_j^0\,\widetilde{\sigma} + \frac{\d G_i}{\d Z_j}\,\widetilde{Z}_j +
\mu\,\frac{\d G_i}{\d \xi_j}\, \d_\theta \widetilde{Z}_j.
\end{align}
Throughout the section we shall refer to the second argument of $G_i$ by the
vector $\xi_i$. Also $D_{ij}'$ denotes the $\sigma$ derivative of $D_{ij}$.
Equation \eqref{rdlin} is rewritten for a single mode $\propto e^{s t}$
so that
\begin{align}
s \widetilde{Z}_i   &= \mathcal{L}_{ij}(\theta)\,\widetilde{Z}_j.
\end{align}
Having introduced these notational conventions, we next make a few preliminary
remarks. 

 Note that the null space of the operator $\mathcal{L}_{ij}$ must
include the function $\d_\theta Z_i^0$. That is to say $\mathcal{L}_{ij}
\d_\theta Z_j^0=0$. This is a consequence of the translational symmetry of the
problem. If $Z_i^0(\theta)$ is a solution to \eqref{rdwt} then so must be
$Z_i^0(\theta+d\theta)$. Upon substituting the latter into \eqref{rdwt}, then
expanding and linearising in the small displacement $d\theta$,
 we arrive at the result. In addition, we assume that the null space of
 $\mathcal{L}_{ij}$ is one-dimensional and spanned by $\d_\theta
 Z_i^0$. 

The linear operator possesses periodic coefficients and thus we can make the
Floquet ansatz: $\widetilde{Z}_j = \hat{Z}_j \,\text{exp}(\text{i} k\theta/\mu)$. This introduces
a new wavenumber parameter $k$ upon which the spectrum of
$\mathcal{L}_{ij}$ smoothly depends. We denote the (discrete) eigenvalues of the operator by
$s_n(k)$, where $n$ is some index and the eigenvalues' functional
dependence on $k$ is explicit. Of the various eigenvalues associated with the
system, we are interested in this section with just one: the diffusive mode which governs
the relaxation of wavenumber, and which is the generalisation of the viscous
instability mode. This eigenvalue is zero when $k=0$, i.e.\ $s(0)=0$, and
as we have seen is proportional to $k^2$. It follows that its eigenfunction
must lie in the null space of $\mathcal{L}_{ij}$ when $k=0$. Thus we set
$\hat{Z}_i=\d_\theta Z_i^0$ when $k=0$. All the other modes of
$\mathcal{L}_{ij}$ are assumed to be stable.

Lastly we denote by $Z^\text{ad}_i$ the nontrivial function in the null space of
the adjoint of $\mathcal{L}_{ij}$. This we normalise so that
$$ \langle \bm{Z}^\text{ad},\, \d_\theta \bm{Z}^0 \rangle=1 $$
where we have defined the inner product above by
$$ \langle \bm{Z}_1,\,\bm{Z}_2 \rangle = \int_0^{2\pi}
\bm{Z}_1\cdot\bm{Z}_2\,d\theta.$$

\subsection{The modulational equation}

We introduce the slow time scale $T$ and long length scale $X$ of Section 6.2 and
make the ansatz
$$ Z_i= Z_i^0( \theta+ \Theta, \mu+ \delta \Theta_X) + \delta^2 \widetilde{Z}_i(\theta,X,T)
$$
where the slow modulation $\Theta(X,T)$ is introduced, and $\delta$ is a small
ordering parameter. This is substituted into \eqref{rxndif} and expanded in
orders of $\delta$. The result is rather messy and we skip most of the
steps. At order $\delta^0$ we obtain Eq.~\eqref{rdwt}, which is satisfied
by construction. At order $\delta^1$ we obtain the $\mu$ derivative of
equation \eqref{rdwt} which is also satisfied. 

At order $\delta^2$ the balance
is
\begin{equation}\label{delta2}
\mathcal{L}_{ij} \widetilde{Z}_j = \Theta_T\,\d_\theta Z^0_i -  A_i(\theta)\,\Theta_{XX} - B_i(\theta) \Theta_X^2,
\end{equation}
for complicated vectors $A_i$ and $B_i$.
We can be assured this equation is solvable if the inner product of the right
side with $\bm{Z}^\text{ad}$ is identically zero. This furnishes the evolution
equation for $\Theta$,
\begin{equation}
\Theta_T  - \langle
\bm{Z}^\text{ad},\,\bm{A} \rangle\, \Theta_X^2 -
\langle\bm{Z}^\text{ad},\,\bm{B}\rangle\,\Theta_{XX} =0.
\end{equation}
What we need to do now is compute the inner products which appear in it. 

The vector $A_i$ may be written as
\begin{align*}
A_i &= \frac{1}{2}\mathcal{L}_{ij} \d^2_{\mu} Z_j^0 + c_g \d_{\theta\mu} Z^0_i +
2\mu D_{ij}\d_{\theta\theta\mu} Z_j^0 + D_{ij}\d_\theta^2 Z_j^0 \\
&\hskip1.5cm +D_{ij}'(\mu^2 \d_{\theta\theta\mu} Z_j^0 + 2\mu \d_\theta^2
Z_j^0)\d_\mu \sigma^0 + \frac{1}{2}\mu^2 D_{ij}'' \d_\theta^2 Z_j^0\,(\d_\mu
\sigma^0)^2 \\
&\hskip1.5cm +\frac{1}{2} \frac{\d^2 G_i}{\d Z_j \d Z_m}\,\d_\mu Z^0_j\,\d_\mu
Z^0_m + \frac{\d^2 G_i}{\d Z_j \d \xi_m}\,\d_\mu Z_j^0\,(\mu
\d_{\theta\mu}Z_m^0+\d_\theta Z_m^0) \\
&\hskip1.5cm +\frac{1}{2}\,\frac{\d^2 G_i}{\d \xi_j \d \xi_m}\,
(\mu\d_{\theta  \mu} Z_j^0+\d_\theta Z_j^0  )(\mu\d_{\theta  \mu}
Z_m^0+\d_\theta Z_m^0)+\frac{\d G_i}{\d \xi_j} \d_{\mu\theta} Z_j^0.
\end{align*}
This expression can be remarkably simplified. By differentiating equation
\eqref{rdwt} twice with respect to $\mu$ we obtain an identity which permits us
to cancel all the terms above save one. We find
\begin{align*}
A_i = -\frac{1}{2}\,\omega''(\mu)\,\d_\theta Z_i^0.
\end{align*}
Therefore,
\begin{align}
 \langle \bm{Z}^\text{ad},\, \bm{A} \rangle = -\frac{1}{2}\omega''(\mu).
\end{align}

The vector $B_i$ may be written as
\begin{align*}
B_i= c_g\,\d_\mu Z_i^0 + D_{ij} \d_\theta Z_j^0 + 2\mu D_{ij}\d_{\mu\theta}
Z_j^0 +\frac{\d G_i}{\d \xi_j}\,\d_\mu Z_j^0.
\end{align*}
The inner product of this expression with $Z_i^\text{ad}$ can also be
substantially simplified. We turn to equation \eqref{rdwt} again and take its
first derivative with respect to $\mu$; we next make the ansatz
$\widetilde{Z}_i= \hat{Z}_j\,\text{exp}(\text{i} k\theta/\mu+ st)$,
substitute this into \eqref{rdlin}, and take its
first and second derivative with respect to $k$ while setting $k=0$. These
three equations yield
three solvability conditions from which we derive
\begin{align}
 \frac{d^2s}{dk^2}(0) = -2 \langle \bm{Z}^{\text{ad}},\, \bm{B} \rangle.
\end{align}
In the process of the proof we have needed 
\begin{align}
\langle \bm{Z}^{\text{ad}},\, \d_\mu \bm{Z}^0 \rangle =0
\end{align}
and the identification $\d_k \hat{Z}_i= \d_\mu Z^0_i$ at $k=0$. These proceed
from the translational invariance of the system.

The final equation for $\Theta$ is
$$ \Theta_T + \frac{1}{2}\omega''(\mu)\,\Theta_X^2 +
\frac{1}{2}s''(0)\,\Theta_{XX}=0,$$
from which we can derive the Burgers equation,
Eq.~\eqref{Burgers}. Also, the previous appendix supplies us with expressions
for $\omega''(\mu)$ and $s''(0)$ to leading order in small $\mu$, and this
allows us to connect the Burgers equation to the material parameters of the
disk. From Eqs \eqref{asomega} and \eqref{modvisc}:
$$ \omega''(\mu)= 2 F_1(q_0), \qquad s''(0)= 2(\beta+1)F_3(q_0),$$
where $q_0$ is the root of $F_2(q)=0$.


\begin{thebibliography}{40}

\bibitem{ArT}
Araki, S., Tremaine, S., 1986. The Dynamics of Dense Particle Disks. 
\emph{Icarus}, \textbf{65}, 83-109.

\bibitem{ArKr}
Aranson, I.~S., Kramer, L., 2002. The world of the complex Ginzberg-Landau
equation.
\emph{Reviews of Modern Physics}, \textbf{74}, 99-143.

\bibitem{Bern88}
Bernoff, A.~J., 1988. Slowly varying fully nonlinear wavetrains in the
Ginzburg-Landau equation. \emph{Physica D}, \textbf{30}, 363-381.



\bibitem{Bord}
Borderies, N., Goldreich, P., Tremaine, S., 1983.
Perturbed particle disks. \emph{Icarus}, \textbf{55}, 124-132.

\bibitem{Bord2}
Borderies, N., Goldreich, P., Tremaine, S., 1985. A granular flow model for
dense planetary rings. \emph{Icarus}, \textbf{63}, 406-420.

\bibitem{Bord3}
Borderies, N., Goldreich, P., Tremaine, S., 1986.
Nonlinear density waves in planetary rings. \emph{Icarus},
\textbf{68}, 522-533.

\bibitem{Brahic}
Brahic, A., 1977. 
Systems of colliding bodies in a gravitational field. I - Numerical simulation of the standard model
\emph{Astronomy and Astrophysics}, \textbf{54}, 895-907.

\bibitem{bridge}
Bridges, F., Hatzes, A., Lin, D.~N.~C., 1984. Structure, Stability and
Evolution of Saturn's Rings. \emph{Nature}, \textbf{309}, 333-335.

\bibitem{Chate}
Chat\'{e}, H., 1994. Spatio-temporal intermittency regimes of the
one-dimensional complex Ginzburg-Landau equation. \emph{Nonlinearity},
\textbf{7}, 185-204.

\bibitem{Col}
Colwell, J.~E., Esposito, L.~W., Sremcevic, M., Stewart, G.~R., McClintock,
        W.~E., 2007. Self-gravity wakes and radial structure of Saturn's B
        ring. \emph{Icarus}, \textbf{190}, 127-144.

\bibitem{Dais}
Daisaka, H., Tanaka, H., Ida, S., 2001. 
Viscosity in a Dense Planetary Ring with Self-Gravitating Particles.
\emph{Icarus}, \textbf{154}, 296-312. 


\bibitem{Doel09}
Doelman, A., Sandstede, B., Scheel, A., Schneider G. 2009.
The dynamics of modulated wave trains.
\emph{Memoirs of the American Mathematical Society} (accepted).

\bibitem{Durisen}
Durisen, R.~H., 1995. An instability in planetary rings due to
ballistic transport. \emph{Icarus}, \textbf{115}, 66-85.

\bibitem{Ei}
Ei, S.-I., 2002. The motion of weakly interacting pulses in reaction--diffusion
systems. \emph{Journal of Dynamics and Differential Equations}, \textbf{14}, 85-137.

\bibitem{Elph}
Elphick, C., Meron, E., Spiegel, E.~A., 1990. Patterns of propagating pulses. \emph{SIAM Journal of Applied  Mathematics}, \textbf{50}, 490-503.

\bibitem{French}
French, R.~G., Salo, H., McGhee, C.~A., Dones, L., 2007. 
HST observations of azimuthal asymmetry in Saturn's rings. \emph{Icarus}, 
\textbf{189}, 493-522.

\bibitem{Fromage}
Fromang, S., Papaloizou, J., 2007. Properties and stability of freely
propagating nonlinear density waves in accretion disks. \emph{Astronomy and
  Astrophysics}, \textbf{468}, 1-18.

\bibitem{GHC}
Gibson, J.~F., Halcrow, J., Cvitanovi\'{c}, P., 2008.
Visualising the geometry of state space in plane Couette flow. \emph{Journal
  of Fluid Mechanics}, \textbf{611}, 107-130.

\bibitem{GL65}
Goldreich, P., Lynden-Bell, D., 1965. II. Spiral arms as sheared gravitational
instabilities. \emph{Monthly Notices of the Royal Astronomical Society},
\textbf{130}, 125-158.

\bibitem{GT78a}
Goldreich, P., Tremaine, S., 1978a.
The excitation and evolution of density waves. \emph{Astrophysical Journal},
\textbf{222}, 850-858.

\bibitem{GT78b}
Goldreich, P., Tremaine, S., 1978b.
The velocity dispersion of Saturn's rings.
\emph{Icarus}, \textbf{34}, 227-239.

\bibitem{GT79}
Goldreich, P., Tremaine, S., 1979.
The excitation of density waves at the Lindblad and corotation resonances by
an external potential. \emph{Astrophysical Journal},
\textbf{233}, 857-871.

\bibitem{GT80}
Goldreich, P., Tremaine, S., 1980.
Disk-satellite interactions. \emph{Astrophysical Journal},
\textbf{241}, 425-441.

\bibitem{Ham4}
H\"{a}meen-Anttila, K.~A., Salo, H., 1993.
Generalised Theory of Impacts in Particulate Systems.
\emph{Earth, Moon, and Planets}, \textbf{62}, 47-84.

\bibitem{HK} 
Howard, L.~N., Kopell, N., 1977.
Slowly varying waves and shock structures in reaction--diffusion equations. 
\emph{Studies in Applied Mathematics}, \textbf{56}, 95-145.


\bibitem{InRow}
Infeld, E., Rowlands, G., 1990. \emph{Nonlinear Waves, solitons and
  chaos}. Cambridge University Press, Cambridge.

\bibitem{Ip}
Ipsen, M., van Hecke, M., 2001. Composite ``zigzag'' structures in the 1D
complex Ginzburg-Landau equation. \emph{Physica D}, \textbf{160}, 103-115.

 \bibitem{K78}
    Kato, S., 1978. Pulsational instability of accretion disks to
    axially symmetric oscillations. \emph{Monthly Notices of the Royal
Astronomical Society}, \textbf{185}, 629-642.

\bibitem{K00}
Kato, S., 2000. Basic Properties of Thin-Disk Oscillations.
\emph{Publications of the Astronomical Society of Japan}, \textbf{53}, 1-24.


\bibitem{LO06a}
Latter, H.~N., Ogilvie, G.~I., 2006a. The linear stability of dilute
particulate rings. \emph{Icarus}, \textbf{184}, 498-516.

\bibitem{LO06b}
Latter, H.~N., Ogilvie, G.~I., 2006b. Viscous overstability and eccentricity
evolution in three-dimensional gaseous discs.
 \emph{Monthly Notices of the Royal
Astronomical Society}, \textbf{372}, 1829-1839.

\bibitem{LO08}
Latter, H.~N., Ogilvie, G.~I., 2008. Dense planetary rings and the viscous overstability.
\emph{Icarus}, \textbf{195}, 725-751.


\bibitem{Lin}
Lin, D.~N.~C., Bodenheimer, P., 1981. On the Stability of Saturn's
Rings.
\emph{The Astrophysical Journal}, \textbf{248}, L83-L86.

\bibitem{LR95}
Longaretti, P.~Y., Rappaport, N., 1995. Viscous Overstabilities in Dense Narrow
Planetary Rings. \emph{Icarus}, \textbf{116}, 376-396.


\bibitem{Lyu}
Lyubarskij, Y.~E., Postnov, K.~A., Prokhorov, M.~.E., 1994. 
Eccentric Accretion Discs. \emph{Monthly Notices of the Royal Astronomical
  Society},
\textbf{266}, 583-596.


\bibitem{Oggie01}
Ogilvie, G.~I., 2001. Non-linear fluid dynamics of eccentric discs.
\emph{Monthly Notices of the Royal Astronomical Society}, \textbf{325}, 231-248.



\bibitem{papalin}
Papaloizou, J.~C.~B., Lin, D.~N.~C., 1988.
On the pulsational overstability in narrowly confined viscous rings.
\emph{Astrophysical Journal}, \textbf{331}, 838-860.

\bibitem{Popp}
Popp, S., Stiller, O., Aranson, I., Weber, A, Kramer, L., 1993. Localized Hole
Solutions and Spatiotemporal Chaos in the 1D Complex Ginzburg-Landau Equation.
\emph{Physical Review Letters}, \textbf{70}, 3880-3883.

\bibitem{Science}
Porco, C.~C. and 34 colleagues, 2005. Cassini Imaging Science: Initial Results
on Saturn's Rings and Small Satellites.  
\emph{Science}, \textbf{307}, 1226-1236.


\bibitem{Rob}
Robinson, J.~C., 2001. \emph{Infinite-Dimensional Dynamical Systems: An Introduction
to Dissipative Parabolic PDEs and Theory of Global
Attractors}. Cambridge University Press, Cambridge, England. 



\bibitem{Salo91}
Salo, H., 1991. Numerical simulations of dense collisional systems.
\emph{Icarus}, \textbf{92},  367-368.


\bibitem{Salo01}
Salo, H., 2001. Numerical Simulations of the Collisional Dynamics of Planetary
Rings. In: P\"{o}schel, T., Luding, S., (eds).  
\emph{Granular Gases}. Lecture Notes in Physics,
\textbf{564}. Springer-Verlag, Berlin, 330.




\bibitem{Sally}
Salo, H., Schmidt, J., Spahn, F., 2001. Viscous Overstability in
Saturn's B Ring: I. Direct Simulations and Measurement of Transport
Coefficients. 
\emph{Icarus}, \textbf{153}, 295-315.

\bibitem{SS07}
Salo, H., Schmidt, J., 2007. Dynamical and Photometric modelling of
dense planetary rings. European Planetary Science Congress, Potsdam, Germany, 2007.

\bibitem{Schmiddie}
Schmidt, J., Salo, H., 2003. Weakly Nonlinear Model for Oscillatory
Instability in Saturn's Dense Rings. \emph{Physical Review Letters},
\textbf{90}, 061102, 1-4.


\bibitem{Schit}
Schmidt, J., Salo, H., Spahn, F., Petzschmann, O., 2001. Viscous
Overstability in Saturn's B-Ring: II. Hydrodynamic Theory and
Comparison to Simulations. \emph{Icarus}, \textbf{153}, 316-331.

\bibitem{Shit}
Schmit, U., Tscharnuter, W.~M., 1995. A Fluid Dynamical Treatment of
the Common Action of Self Gravitation, Collisions and Rotation in
Saturn's B-Ring. \emph{Icarus}, \textbf{115}, 304-319.

\bibitem{Shit2}
Schmit, U., Tscharnuter, W.~M., 1999. On the Formation of the
Fine-Scale Structure in Saturn's B Ring. \emph{Icarus} \textbf{138}, 173-187.


\bibitem{Spa}
Spahn. F., Schmidt, J., 2006. Hydrodynamic Description of Planetary Rings.
\emph{GAMM-Mitteilungen}, \textbf{29}, 115-140 .

\bibitem{Spahn}
Spahn, F., Schmidt, J., Petzschmann, O., 2000. Stability Analysis of
a Keplerian Disk of Granular Grains: Influence of Thermal Diffusion. 
\emph{Icarus}, \textbf{145}, 657-660.

\bibitem{Shr}
Shraiman, B.~I., Pumir, A., van Saarloos, W., Hohenberg, P.~C., Chat\'{e}, H.,
Holen, M., 1992. Spatiotemporal chaos in the one-dimensional complex Ginzburg-Landau
equation. \emph{Physica D}, \textbf{57}, 241-248.



\bibitem{SYL85}
Shu, F.~H., Yuan, C., Lissauer, J.~J., 1985a.
Nonlinear spiral density waves: an inviscid theory.
\emph{The Astrophysical Journal}, \textbf{291}, 356-376.

\bibitem{SDLYC85}
Shu, F.~H., Dones, L., Lissauer, J.~J., Yuan, C., Cuzzi, J.~N., 1985b.
Nonlinear spiral density waves: viscous damping.
\emph{The Astrophysical Journal}, \textbf{299}, 542-573.



\bibitem{SLB}
Stewart, G.~R., Lin, D.~N.~C., Bodenheimer, P., 1984. Collisional
Transport Processes. In: Greenberg, R., Brahic, A.
(eds). \emph{Planetary Rings}. University of Arizona Press, Tucson, Arizona.

\bibitem{TMTFR07}
Thomson, F~S., Marouf, E.~A., Tyler, G.~L., French, R.~G., Rappoport, N.~J., 2007.
Periodic microstructure in Saturn's rings A and B. Geophysical Research
Letters, \textbf{34}, \textbf{24}, L24203.

\bibitem{tre}
Tremaine, S., 2003. On the Origin of Irregular Structure in Saturn's
Rings. \emph{The Astronomical Journal}, \textbf{125}, 894-901.

\bibitem{Ward}
Ward, W.~R., 1981. On the Radial Structure of Saturn's Rings. 
\emph{Geophysical Research Letters}, \textbf{8}, 641-643.

\bibitem{Whit}
Whitham, G.~B., 1974. \emph{Linear and Nonlinear Waves}. John Wiley, New York.


\bibitem{WT}
Wisdom, J., Tremaine, S., 1988. Local Simulations of Planetary Rings. 
\emph{The Astronomical Journal}, \textbf{95}, 925-940.

\end{thebibliography}
\end{document}